\documentclass[12pt]{article}

\usepackage[title]{appendix}
\usepackage[T1]{fontenc}
\usepackage{ae,aecompl}
\usepackage[dvips]{graphicx}
\usepackage{amssymb}
\usepackage{amsmath}
\usepackage{subfigure}
\usepackage[round,authoryear]{natbib}
\usepackage{color}
\usepackage{array}
\usepackage{rotating}
\usepackage{colortbl}
\usepackage{tikz}
\usetikzlibrary{patterns}

\usepackage[normalem]{ulem}

\usepackage[margin=1in]{geometry}

\definecolor{webgreen}{rgb}{0,0.4,0}
\definecolor{webbrown}{rgb}{0.6,0,0}
\definecolor{purple}{rgb}{0.5,0,0.25}
\definecolor{darkblue}{rgb}{0,0,0.7}
\definecolor{darkred}{rgb}{0.7,0,0}
\definecolor{darkgreen}{rgb}{0,0.7,0}
\usepackage[pdfborder=false]{hyperref}
\hypersetup{colorlinks,citecolor=darkred,filecolor=black,linkcolor=darkblue,urlcolor=webgreen,pdfpagemode=none,
pdfstartview=FitH}
\newcommand{\ignore}[1]{}
\newtheorem{lemma}{{\sc Lemma}}
\newtheorem{remark}{{\sc Remark}}
\newtheorem{prop}{{\sc Proposition}}
\newtheorem{cor}{{\sc Corollary}}
\newtheorem{theorem}{{\sc Theorem}}
\newtheorem{defn}{{\sc Definition}}

\newtheorem{claim}{{\sc Claim}}
\newtheorem{example}{{\sc Example}}

\usepackage{sectsty}
\sectionfont{\centering\normalfont\scshape}
\subsectionfont{\centering\normalfont}

\newcommand{\vlb}{{\underline{v}}}
\newcommand{\vub}{{\overline{v}}}

\newcommand{\tinyh}{\mbox{\tiny H}}
\newcommand{\tinyi}{\mbox{\tiny I}}
\newcommand{\tinym}{\mbox{\tiny M}}

\allowdisplaybreaks

\begin{document}

\begin{titlepage}
\title{\textbf{\large{Rank-preserving Multidimensional Mechanisms:\\
an equivalence between identical-object and heterogeneous-object models}}\thanks{We are grateful 
to Gabriel Carroll, Rahul Deb, Gagan Ghosh, Alessandro Pavan, Takuro Yamashita and to seminar participants at Academia Sinica, Ashoka University, Bar-Ilan University, UCLA, IIM Bangalore, 2021 Conference of the Society for 
Advancement of Economic Theory (SAET), 2021 BRICS NU Conference, and 2022 Conference 
on Mechanism Design, Singapore for thoughtful comments. We are also grateful to three anonymous referees and an associate editor for their valuable feedback and suggestions. Debasis Mishra acknowledges 
financial support from the Science and Engineering Research Board (SERB Grant No. SERB/CRG/2021/003099) of India.}}
\author{Sushil Bikhchandani\thanks{Anderson School at UCLA, Los Angeles 
({\tt sbikhcha@anderson.ucla.edu}).}$\;\;$ and 
Debasis Mishra\thanks{Indian Statistical Institute, Delhi ({\tt dmishra@isid.ac.in}).} \\
}

\maketitle

\begin{abstract}

	We show that the mechanism-design problem for a monopolist selling multiple, 
	heterogeneous objects to a buyer with ex ante symmetric and additive values is 
	equivalent to the mechanism-design problem for a monopolist selling identical objects to a buyer
	with decreasing marginal values. We derive three new results for the identical-objects model: (i) a new condition for revenue monotonicity of stochastic mechanisms, (ii) a sufficient condition on priors, such that prices in optimal deterministic mechanism are not increasing, and (iii) a simplification of incentive constraints for deterministic mechanisms. We use the equivalence to establish corresponding results in the heterogeneous-objects model.
	
	\bigskip
	
	\medskip
	
	
	\end{abstract}
	
	\thispagestyle{empty}
	\end{titlepage}
	
	\section{Introduction}

	We study two models for selling multiple, indivisible objects to a buyer: a heterogeneous-objects model and an identical-objects model. The seller chooses an incentive compatible (IC) and individually rational (IR) mechanism with the goal of maximizing expected revenue. The buyer's type is multidimensional, a vector consisting of (marginal) values for each object. With heterogeneous objects, the buyer's value for a bundle of objects is the sum of the values of objects in the bundle. With identical objects, the buyer's value for $k$ units of the object is the sum of the marginal values of these units. The buyer's values are private information. The seller knows the distribution of buyer values.
	
	We show that any heterogeneous-objects model with decreasing marginal values is equivalent to an identical-objects model in the following sense. There is a bijective mapping between the set of symmetric,\footnote{A mechanism is {\sl symmetric} if a permutation of the allocation probabilities (of objects) at a buyer type is equal to the allocation probabilities at the same permutation of the buyer type.} IC and IR mechanisms in the heterogeneous-objects model and the set of IC and IR mechanisms in the identical-objects model. If the distribution of buyer values in the heterogeneous-objects model is exchangeable\footnote{A distribution is exchangeable if it is symmetric.} then (i) the expected revenue of a symmetric, IC and IR mechanism in the heterogeneous-objects model is equal to the expected revenue of its equivalent mechanism in the identical-objects model (when the distribution of buyer values is a projection of the distribution in the heterogeneous-objects model) and (ii)~in the heterogeneous-objects model, there exists an optimal mechanism that is symmetric. Hence, the optimal revenues in the two models are equal.

	While exchangeability is a strong assumption as it entails a presumption of ex ante symmetric buyer values, 
it is plausible when the seller is uninformed about buyer preferences. Moreover, exchangeability is a weaker assumption than  i.i.d.\,distribution of buyer values, which is often assumed in the literature. 
	
	With $n$ heterogeneous objects, allocation rules  induce probability distributions with $2^n$ outcomes (bundles of objects) whereas with $n$ identical objects, allocation rules induce probability distributions with only $n+1$ outcomes (number of objects). Therefore, as it has a smaller allocation space, the identical-objects model is a more tractable setting than the heterogeneous-objects model  for the discovery of new results (as we demonstrate in our applications). These results can be adapted to the exchangeable, heterogeneous-objects model via the equivalence. The equivalence is also useful in adapting known results for heterogeneous objects to identical objects.
	
	We provide three applications to demonstrate the usefulness of the equivalence. In these applications, we establish novel results in the identical-objects model and use the equivalence to establish new results in the heterogeneous-objects model.
	
	First, we obtain results on revenue monotonicity. \citet{HR15} show that the optimal revenue need not be monotone in the distribution of values. We obtain a new sufficient condition, {\sl majorization monotonicity}, for a mechanism to be revenue monotone in the identical-objects model and therefore also in the heterogeneous-objects model with an exchangeable distribution. If an optimal mechanism is symmetric and almost deterministic (i.e., there is randomization over at most one outcome in the range of the mechanism), then it satisfies majorization monotonicity; consequently, the optimal revenue is monotone.
	
	In a second application, we show that in the identical-objects model, optimal prices in the class of deterministic mechanisms cannot be increasing if the marginal distributions of values satisfy the hazard-rate order.\footnote{We show in an example that despite decreasing marginal values, optimal prices may be increasing in the identical-objects model even though marginal values are decreasing.} The equivalence implies that in the heterogeneous-objects model with i.i.d. distribution of values, optimal prices in the class of symmetric and deterministic mechanisms cannot be supermodular. Thus, when two objects are for sale, optimal deterministic mechanisms are submodular. 
	
	In a third application, we show that in the identical-objects model a weaker notion of incentive compatibility implies full incentive compatibility for deterministic mechanisms. This weaker notion, which we call {\sl upper-set IC} or UIC, requires incentive constraints only for pairs $(v,v')$ such that $v \ge v'$ or $v' \ge v$.  Under a mild condition,\footnote{This condition, which we call {\sl object non-bossiness,} requires that if the allocation of object~$i$ is the same for types $(v_i,v_{-i})$ and $(v'_i,v_{-i})$, then the entire allocation vector to these two types is the same. We show that object non-bossiness is generically satisfied by any IC mechanism.} every deterministic UIC mechanism is IC in the identical-objects model. Therefore, every symmetric, deterministic, rank-preserving, UIC mechanism is IC in the heterogeneous-objects model. This leads to a simplification of the design problem for deterministic mechanisms, which we plan to investigate in future research.
	
\medskip

	To establish the equivalence between the two models, it is straightforward to show that any IC and IR mechanism in the identical-objects model can be extended to a symmetric mechanism in the heterogeneous-objects model, while preserving IC and IR. In the other direction, a complication is that the restriction of an IC and IR  mechanism in the heterogeneous-objects model to the domain of identical objects need not yield a mechanism that is feasible in the identical-objects model. This is because in order to allocate the  $(i+1)^{\mbox{\footnotesize st}}$ unit in the identical-objects model, the $i^{\mbox{\footnotesize th}}$ unit must also be allocated; thus, feasibility requires that the $i^{\mbox{\footnotesize th}}$ unit is allocated with a (weakly) greater probability than the $(i+1)^{\mbox{\footnotesize st}}$ unit. There is no such feasibility restriction in the heterogeneous-objects model.
	
	The property of rank preserving plays a key role in showing that a symmetric mechanism in the heterogeneous-objects model maps into a feasible mechanism in the identical-objects model.  A mechanism is {\sl rank preserving} if whenever the (buyer's) value for object~$i$ is greater than the value for object~$j$, the probability that object $i$ is allocated to the buyer is at least as large as the probability that object~$j$ is allocated. 
	 
	 In the identical-objects model, decreasing marginal values imply that any feasible mechanism is rank preserving. In the heterogeneous-objects case, there exist feasible, IC and IR mechanisms that are not rank preserving; however, if a symmetric mechanism is IC, then we show that it must be rank preserving. This is critical in establishing equivalence.
	
	Under exchangeability, in the heterogeneous-objects model there exists an optimal mechanism that is symmetric. Therefore, it must be rank-preserving and has an equivalent mechanism in the identical-objects model; this equivalent mechanism must be optimal in the identical-objects model.
	
	A general solution to the optimal mechanism-design problem for the sale of multiple objects is unknown. Much of the multidimensional screening literature has focused on the sale of heterogeneous objects (see, for instance, \cite{MM88}, \cite{T04}, \cite{MV06}, \cite{HR15}, \cite{Ca17}, \cite{DDT17}). The few papers on multidimensional screening that focus on the sale of  identical objects do so in models with two-dimensional private information. In \cite{MV09} and \cite{DHP20}, a buyer has the same value for each unit of the object up to a capacity, after which the value for additonal units is zero. Both the value and the capacity of a buyer are private information.  \cite{DLN12} consider budget-constrained buyers who have the same value for each unit. They obtain impossibility results regarding achieving efficiency when each buyer's value and budget are private information. \cite{BM22} provide sufficient conditions on priors for an optimal mechanism for the sale of two identical objects to be deterministic and almost deterministic. None of these papers consider $n\ge 3$ dimensional private information, which is a challenging setting.

	The paper is organized as follows. We present the heterogeneous-objects model in Section~\ref{se:mdl} and establish the connection between rank preserving and symmetry in Section~\ref{se:srpm}. The identical-objects model and the equivalence between the two models is presented in Section~\ref{se:eql}. The three applications are in Section~\ref{se:apps}. We end with a discussion in Section~\ref{sec:disc}.  All the proofs are in an appendix, including some in an online appendix.

\section{The heterogeneous-objects model}\label{se:mdl}

There is a set of heterogeneous objects denoted by $N=\{1,2,\ldots,n\}$. The type of the buyer is a vector of valuations $v := (v_1,v_2,\ldots,v_n)$, where each $v_i\in [\vlb,\vub]$, $0\le \vlb<\vub<\infty$. The buyer type space is \vspace{-2mm}
\begin{align*}
\overline{D}^{\tinyh}:=[\vlb,\vub]^n
\end{align*}
For any type $v$, $v_i$ is the buyer's value for object~$i$ and the value for a bundle of objects $S \subseteq N$ is additive: $\sum_{i \in S}v_i$. As  the $n$ objects may be distinct, there is no restriction on values across objects, i.e., both $v_i > v_j$ or $v_j < v_i$ are possible. The values $v_1,v_2,\ldots,v_n$ are jointly distributed with cumulative distribution function (cdf) $F^{\tinyh}$ and density function $f^{\tinyh}$ with support $\overline{D}^{\tinyh}$. An alternate interpretation is that there is a unit mass of buyers distributed with density~$f^{\tinyh}$. 

The seller is the mechanism designer. We assume that the seller's cost for each object is not more than $\underline{v}$. Hence, without loss of generality, we normalize costs to zero. We refer to the heterogeneous-objects model as $\mathcal{M}^{\tinyh} :=(N, \overline{D}^{\tinyh}, f^{\tinyh})$. 

A {\bf mechanism}  is an allocation probability vector $q:\overline{D}^{\tinyh} \rightarrow [0,1]^n$ and a payment $t:\overline{D}^{\tinyh} \rightarrow \Re$. A buyer with (reported) type $v$ is allocated object~$i$ with probability $q_i(v)$, $i=1,2,\ldots, n$ and makes a payment of $t(v)$. 
The expected utility
of a buyer of type $v$ from mechanism $(q,t)$ is
\begin{align*}
u(v):= v \cdot q(v) - t(v)
\end{align*}

A mechanism $(q,t)$ is {\bf deterministic} if $q_i(v) \in \{0,1\}$ for all $v$ and all $i$. A mechanism that is not deterministic is a {\bf stochastic} mechanism.

A mechanism $(q,t)$ is {\bf incentive compatible (IC)} if for every $v,v' \in \overline{D}^{\tinyh}$, we have
\begin{align*}
u(v) &\ge v \cdot q(v') - t(v') = u(v') + (v-v')\cdot q(v')
\end{align*}
A mechanism $(q,t)$ is {\bf individually rational (IR)} if for every $v \in \overline{D}^{\tinyh}$, $u(v) \ge 0$. If $(q,t)$ is IC, then it is IR if and only if $u(\underline{v},\ldots,\underline{v}) \ge 0$.

We restrict attention to mechanisms $(q,t)$ that satisfy $u(\underline{v},\ldots,\underline{v})=0$.  This is without
loss of generality as the seller is interested in maximizing expected revenue. Then IC implies
that $0=u(\underline{v},\ldots,\underline{v}) \ge \underline{v} \big[\sum_i q_i(v)\big] - t(v)$ or $t(v) \ge \underline{v} \big[\sum_i q_i(v)\big]$. Note that
IR implies $t(v) \le v \cdot q(v)$. Since the domain of types is bounded, this implies that for any IC and IR mechanism $(q,t)$, both $u(v)$ and $t(v)$ are bounded above and below for every $v$.

\subsection{Symmetric and Rank-preserving Mechanisms}\label{se:srpm}

We formally define symmetric and rank-preserving IC mechanisms in model $\mathcal{M}^{\tinyh}$, and show that these properties are closely related. Before we do so, we restrict attention to strict types without loss of generality.
A type vector $v$
is {\bf strict} if $v_i \ne v_j$ for all $i,j \in N$. Let $D^{\tinyh}$ denote
the {\bf set of all strict types} in $\overline D^{\tinyh}$.

\begin{lemma}\label{le:int}
Let $(q,t)$ be an IC and IR mechanism defined on $D^{\tinyh}$. There exists an IC and IR mechanism $(\bar q, \bar t)$ defined on $\overline{D}^{\tinyh}$ such that
\begin{align*}
(\bar q(v), \bar t(v)) = ( q(v),  t(v)) \qquad \forall v\in D^{\tinyh}
\end{align*}
\end{lemma}

Throughout, we assume that the probability distribution of types has a density function. Thus, the set of non-strict types has zero probability. Consequently, the expected revenue from $(\bar q, \bar t)$ is the same as the expected revenue from $(q,t)$. Hence, Lemma~\ref{le:int} allows us to define mechanisms on the set of strict types, i.e., on $D^{\tinyh}$, and then extend them to $\overline{D}^{\tinyh}$. This results in a simplification of the proofs. 

Let $\sigma$ represent a permutation of the set $N$. The identity permutation is $\sigma^{\tinyi} := (1,\ldots,n)$. The set of all permutations of $N$ is denoted by $\Sigma$. We partition the set of strict types, $D^{\tinyh}$, using permutations in $\Sigma$. For any permutation $\sigma \in \Sigma$, let
\begin{align}\label{eq:sigma}
D({\sigma}) & = \{v \in D^{\tinyh}: v_{\sigma(1)} > v_{\sigma(2)} > \ldots > v_{\sigma(n)}\}
\end{align}
Note that $D^{\tinyh} \equiv \cup_{\sigma \in \Sigma}D({\sigma})$ and $D(\sigma)\cap D(\sigma')=\emptyset$ if $\sigma \neq \sigma'$.

Every type in $D^{\tinyh}$ can be mapped to a type in $D(\sigma^{\tinyi})$. To see this, take any $v\in D^{\tinyh}$. There exists a unique $\sigma$ such that $v\in D(\sigma)\subset D^{\tinyh}$. Let $v^{\sigma}$ denote the permuted type of $v$, i.e., $v^{\sigma}_j=v_{\sigma(j)}$ for all $j \in N$. Eq. (\ref{eq:sigma}) implies that $v^{\sigma}\in D(\sigma^{\tinyi})$. Thus, for an arbitrary type $v \in D^{\tinyh}$ and a permutation $\sigma$, $v^\sigma\in D(\sigma^{\tinyi})$  if and only if $v\in D(\sigma)$.

We start with a mechanism defined on $\overline D^{\tinyh}$ and assume that it satisfies the properties of symmetry and rank preserving, defined below, on the subset $D^{\tinyh}$. As $\overline D^{\tinyh}\backslash D^{\tinyh}$ has zero measure, these properties are satisfied for almost all $v \in \overline D^{\tinyh}$.

\begin{defn}
A mechanism $(q,t)$ is {\bf symmetric} if for every $v \in D^{\tinyh}$ and for every $\sigma \in \Sigma$,\vspace{-3mm}
\begin{align*}
q_i(v^{\sigma}) &=  q_{\sigma(i)}(v)~\qquad~\forall~i \in N \\
t(v^{\sigma}) &= t(v)
\end{align*}
\end{defn}

In a symmetric mechanism, the allocation probabilities at a permutation of type $v$ are the permutation of allocation probabilities at $v$, while the payment function is invariant to permutations of $v$.\footnote{We do not impose symmetry on non-strict types as it implies additional restrictions. To see this,
consider $n=4$ and suppose that $v \equiv (1,0,0,1)$ and $v' \equiv (0,1,1,0)$.
Type $v'$ can be obtained from $v$ via two permutations: $\sigma(1)=2,\sigma(2)=1,\sigma(3)=4,
\sigma(4)=3$ and $\sigma'(1)=3,\sigma'(2)=4,\sigma'(3)=1,\sigma'(4)=2$. Hence, $v'=v^{\sigma}=v^{\sigma'}$.
Applying symmetry here would mean $q_i(v^{\sigma})=q_{\sigma(i)}(v)=q_i(v^{\sigma'})=q_{\sigma'(i)}(v)$ 
for all~$i$ which implies $q_2(v)=q_3(v)$ and $q_1(v)=q_4(v)$. Such restrictions limit the set of 
mechanisms that can be considered. } Later, we show in Proposition~\ref{prop:sym2} that in an exchangeable environment, it is without loss of generality to consider symmetric mechanisms.

To construct a symmetric mechanism, it is enough to define the mechanism
on $D(\sigma^{\tinyi})$, say, and then extend it to $D^{\tinyh}$ symmetrically (as made precise later in Definition~\ref{de:sym} and Lemma~\ref{le:symex}). The following property plays a crucial role in maintaining incentive compatibility in such symmetric extensions.

\begin{defn}
A mechanism $(q,t)$ is {\bf rank preserving} if for every $v \in D^{\tinyh}$ and every $i,j$,
we have $q_i(v) \ge q_j(v)$ if $v_i > v_j$.
\end{defn}

An IC mechanism need not be rank preserving. For example, a  mechanism that allocates object 1 and no other object for zero payment to all types is IC and IR, but not rank preserving. This mechanism is not symmetric. Thus, asymmetric, IC, and IR mechanisms need not be rank preserving.

As shown next, if a symmetric mechanism is IC then it must be rank preserving. Conversely, if a symmetric mechanism is rank preserving and IC on $D(\sigma^{\tinyi})$, then it is IC on $D^{\tinyh}$.

\begin{theorem}
\label{theo:sym0}
Suppose that $(q,t)$ is a symmetric mechanism in $\mathcal{M}^{\tinyh}$. Then, the following are equivalent:\vspace{-3mm}
\begin{enumerate}
\item[(i)] $(q,t)$ is IC on $D^{\tinyh}$.\vspace{-3mm}
\item[(ii)] $(q,t)$  is rank preserving and $(q,t)$ restricted to $D(\sigma^{\tinyi})$ is IC.
\end{enumerate}
\end{theorem}

As noted earlier, a mechanism that allocates object~1 for zero payment to all types is asymmetric but not rank preserving. Thus symmetry is essential for the direction $(i)\implies (ii)$ in Theorem~\ref{theo:sym0}. Symmetry is also essential for the direction $(ii)\implies (i)$. To see this, consider two objects and type space $D^{\tinyh}$ consisting of strict types in $[0,1]^2$. A mechanism sells object~$1$ for free in the lower triangle $D(\sigma^1):=\{(v_1,v_2): v_1 > v_2\}$ and does not sell any objects in the upper triangle $D(\sigma^2):=\{(v_1,v_2): v_1 < v_2\}$. This is a rank-preserving mechanism which satisfies IC constraints in each $D(\sigma^i)$. But it is clearly not an IC mechanism for $D^{\tinyh}$. Notice that this mechanism is not symmetric.

The following example shows that rank preserving is also essential for Theorem~\ref{theo:sym0}. 

\begin{example}\label{ex:nrp}
{\em Consider two objects with the buyer's valuation $v=(v_1,v_2)$ distributed on the unit square. Let $t(\cdot)\equiv 0$ and
\begin{align*}
q(v_1,v_2) & =
\begin{cases}
(0,1), & \mbox{if }\  v_1>v_2  \\
(1,0), & \mbox{if  }\   v_1<v_2
\end{cases}
\end{align*}
This mechanism is symmetric but not rank preserving. Restricted to $D(\sigma^1)$, this mechanism is IC. It is also IC when restricted to $D(\sigma^2)$. But the mechanism is not IC on $D^{\tinyh}$ as any type in $D(\sigma^1)$ benefits by reporting a type in $D(\sigma^2)$ and vice versa. \hfill $\blacksquare$ }
\end{example}

Is every rank preserving and IC mechanism symmetric? The answer is no as the following example
illustrates.

\begin{figure}[!hbt]
\centering
\includegraphics[width=2.6in]{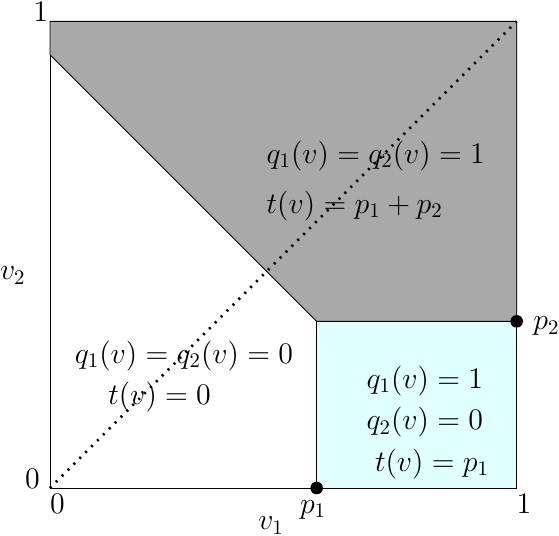}
\caption{A rank-preserving IC mechanism that is not symmetric}
\label{fig:rpasym}
\end{figure}

\begin{example}
\label{ex:rpasym}
{\em Suppose $n=2$ and the type space is $[0,1]^2$. Figure \ref{fig:rpasym} describes
a deterministic mechanism $(q,t)$ for this type space. This mechanism
is clearly IC and rank preserving. But it is not symmetric. \hfill $\blacksquare$} 
\end{example}

We show next that  if the distribution of values is exchangeable, then there exists an optimal mechanism which is rank preserving.

\subsection{Existence of a Rank-preserving Optimal Mechanism}

The expected revenue from an IC and IR mechanism in model $\mathcal{M}^{\tinyh}$ is
\begin{align*}
\textsc{Rev}(q,t;f^{\tinyh}) &:= \int \limits_{\overline{D}^{\tinyh}} t(v)f^{\tinyh}(v)dv = \int \limits_{D^{\tinyh}} t(v)f^{\tinyh}(v)dv
\end{align*}
Further,
\begin{align} \nonumber
\textsc{Rev}(q,t;f^{\tinyh}) &= \sum_{\sigma\in\Sigma} \textsc{Rev}^\sigma(q,t;f^{\tinyh})  \\[5pt] \label{eq:revsig}
\mbox{where}\qquad \textsc{Rev}^{\sigma}(q,t;f^{\tinyh}) & := \int \limits_{D(\sigma)} t(v)f^{\tinyh}(v)dv
\end{align}
We sometimes write $\textsc{Rev}(q,t;F^{\tinyh})$ instead of $\textsc{Rev}(q,t;f^{\tinyh})$.

A mechanism $(q^*,t^*) $ is  {\bf optimal} for density function $f^{\tinyh}$ if it is IC and IR and for any other IC and IR mechanism $(q,t)$\vspace{-2mm}
\begin{align*}
\textsc{Rev}(q^*,t^*;f^{\tinyh}) &\ge \textsc{Rev}(q,t;f^{\tinyh})
\end{align*}
The joint density of values, $f^{\tinyh}$, 
is {\bf exchangeable}\footnote{Strictly speaking, the random variables $v_1, v_2, \ldots, v_n$ are exchangeable.} if
\begin{align*}
f^{\tinyh}(v) &= f^{\tinyh}(v^{\sigma}) \qquad~\forall~v \in D^{\tinyh},\ \forall\sigma \in \Sigma
\end{align*}
Exchangeability is satisfied if $v_1,v_2,\ldots, v_n$ are i.i.d. 
Exchangeable random variables may be positively correlated such as when 
$v_1,v_2,\ldots, v_n$ are distributed i.i.d. conditional on an underlying 
state variable.\footnote{As an example, consider an entity 
that sells ``permits'' for operating in $n$ markets that are ex-ante 
identical. The seller might be a local government that issues licenses for 
liquor stores or a franchisor introducing its product in a new market via 
franchises. The buyer is knowledgeable about market conditions in the $n$ 
markets. The value of market $i$ to the buyer is $v_i=\eta\, m_i$, where 
$\eta$ is the buyer's efficiency level and $m_i$ is the size of market $i$. 
The buyer knows $\eta$  and $m_i$.  The seller has a distribution over $\eta$ 
and has i.i.d. distributions over $m_i$. The random variables $v_i$ are 
exchangeable from the seller's perspective.} Exchangeability also allows for negative correlation, and when $n=2$ this negative correlation can be arbitrarily large.

As shown next, an exchangeable distribution of buyer types in model $\mathcal{M}^{\tinyh}$ allows one to restrict attention to mechanisms that are symmetric and, therefore, also rank preserving.

\begin{prop}\label{prop:sym2}
Suppose that $f^{\tinyh}$ in model $\mathcal{M}^{\tinyh}$ is exchangeable. Then, there exists an optimal mechanism which is symmetric and rank preserving.
\end{prop}

In an exchangeable environment with two-dimensional types, \citet{Pa11a} notes (in the 
proof of his Corollary 1) that an optimal mechanism that is symmetric must exist.\footnote{\citet{MR84} make a similar observation in a single-object auction setting with ex ante symmetric bidders and 
one-dimensional types.} For the sake of completeness, we provide a proof for $n$-dimensional types in an online  appendix (Appendix~\ref{sec:onap}). The proof uses the fact that in the heterogeneous-objects model the simple average of all permutations of an asymmetric, IC and IR mechanism is a symmetric, IC and IR mechanism. Consequently, linearity of the revenue functional and exchangeability imply that for every asymmetric mechanism there exists a symmetric mechanism with the same expected revenue. Thus, in the heterogeneous-objects model, there exists an optimal mechanism that is symmetric and, by Theorem~\ref{theo:sym0}, rank-preserving.

\section{An Equivalence Between Two Models}\label{se:eql}

We now introduce our second model. In the {\bf identical-objects model}, all $n$ objects are identical. The type of the buyer is a vector of marginal valuations $v := (v_1,v_2,\ldots,v_n)$, where each $v_i\in [\vlb,\vub]$, $0\le \vlb<\vub<\infty$. For any type $v$, $v_i$ is the (marginal) value of consuming the $i^{\mbox{\footnotesize th}}$ unit of the object. The value from consuming $k$ units is $\sum_{i=1}^k v_i$. We assume that marginal values are decreasing. Thus, the type space is
\begin{align*}
\overline{D}^{\tinyi} := \big\{v\in  [\vlb,\vub]^n\, \big|\, v_1\ge v_2\ge \ldots\ge v_n\,\big\}
\end{align*}
The values $v_1,\ldots,v_n$ are jointly distributed with cdf $F^{\tinyi}$ and density function $f^{\tinyi}$ with support $\overline{D}^{\tinyi}$.  
The identical-objects model is denoted by $\mathcal{M}^{\tinyi} :=(N, \overline{D}^{\tinyi},f^{\tinyi})$.

A {\bf mechanism} in model $\mathcal{M}^{\tinyi}$ is an allocation probability vector $q:\overline{D}^{\tinyi} \rightarrow [0,1]^n$ and a payment $t:\overline{D}^{\tinyi} \rightarrow \Re$. A buyer with (reported) type $v$ is allocated unit~$i$ with probability $q_i(v)$, $i=1,2,\ldots, n$ and makes a payment of $t(v)$.
In the identical-objects model, $q_i(v)$ denotes the probability of getting the $i^{\mbox{\footnotesize th}}$ unit of the object, which happens whenever $i$ or more units are allocated. In other words, the $(i+1)^{\mbox{\footnotesize st}}$ unit can be allocated only if the $i^{\mbox{\footnotesize th}}$ is also allocated. Thus, 
\begin{align}\label{eq:idrp}
q_i(v)\ge q_{i+1}(v)\qquad\forall v\in \overline{D}^{\tinyi},\  \forall i \in \{1,\ldots,n-1\}
\end{align}
is a feasibility requirement on the allocation probabilities of a mechanism in the identical-objects model.\footnote{If we denote the probability of getting exactly $k$ units by $Q_k(v)$, then $q_i(v)=\sum_{k=i}^nQ_k(v)$. This immediately shows that 
$q_i(v) = Q_i(v)+q_{i+1}(v) \ge q_{i+1}(v)$.}  Feasibility does not impose such a restriction in the heterogeneous-objects model.

The expected utility of a buyer of type $v$ from mechanism $(q,t)$ is
\begin{align*}
u(v):= v \cdot q(v) - t(v)
\end{align*}
The definitions of an IC, IR, and optimal mechanism are similar to those in model $\mathcal{M}^{\tinyh}$. Further, Lemma \ref{le:int} extends to model $\mathcal{M}^{\tinyi}$ so we restrict attention to strict types in $D^{\tinyi}$, and then later extend it to $\overline{D}^{\tinyi}$. Note that $D^{\tinyi}=D(\sigma^{\tinyi})$.

Note that in the identical-objects model, any feasible mechanism is rank preserving. To see this, note that for any $v\in D^{\tinyi}$,  we have $v_i>v_{i+1}$ and by (\ref{eq:idrp}) we have $q_i(v)\ge q_{i+1}(v)$. 
This observation and Theorem \ref{theo:sym0} allow us to establish a formal equivalence 
between the identical-objects and the heterogeneous-objects models.
To this end, we define how a mechanism on $D(\sigma^{\tinyi})$ may be extended symmetrically to $D^{\tinyh}$.
Recall that for every $v\in D(\sigma)$, we have $v^\sigma\in D(\sigma^{\tinyi})$.

\begin{defn}\label{de:sym}
Let $(q,t)$ be a mechanism defined on $D(\sigma^{\tinyi})$ (equivalently on $D^{\tinyi}$).  The {\bf symmetric extension} of $(q,t)$ is a mechanism $(q^s,t^s)$ on $D^{\tinyh}$ such that for every $v \in D(\sigma)$  and for every $\sigma \in \Sigma$\vspace{-3mm}
\begin{align*}
q_{\sigma(i)}^s(v) & = q_i(v^\sigma) \qquad\forall i \\
t^s(v) & = t(v^\sigma)
\end{align*}
\end{defn}

A mechanism defined on $D(\sigma)$, where $\sigma\neq \sigma^{\tinyi}$, may also be extended symmetrically using Definition~\ref{de:sym} after first relabeling the axes. The symmetric extension of a rank-preserving mechanism is IC and IR, as shown next.\footnote{Theorem~\ref{theo:sym0} implies that the symmetric extension of a non-rank-preserving mechanism on $D^{\tinyi}$ will not be IC on $D^{\tinyh}$. Example~\ref{ex:nrp} illustrates this.}

\begin{lemma}\label{le:symex}
Let $(q,t)$ be a rank-preserving, IC and IR mechanism defined on $D(\sigma)$. Then the symmetric extension of $(q,t)$ to $D^{\tinyh}$ is a rank-preserving, IC and IR mechanism.
\end{lemma}

This leads to an equivalence between identical-object models and heterogeneous-object models:

\begin{theorem}\label{theo:sym} $\ \ $\\ \vspace{-10mm}
\begin{itemize}
\item[(i)] Any IC and IR mechanism in model $\mathcal{M}^{\tinyi}$ 
can be extended to a symmetric, IC, and IR mechanism in model 
$\mathcal{M}^{\tinyh}$. \vspace{-2mm}
\item[(ii)] The restriction of any symmetric,
IC, and IR mechanism in model $\mathcal{M}^{\tinyh}$ to $D(\sigma^{\tinyi})$ defines an IC and IR mechanism 
in model $\mathcal{M}^{\tinyi}$.\vspace{-2mm}
\item[(iii)] If the density $f^{\tinyh}$ is exchangeable with $f^{\tinyi}(v) = n! f^{\tinyh}(v)~\forall~v \in \overline{D}^{\tinyi}$ then \vspace{-2mm}
\begin{itemize}
\item[(a)] an IC and IR mechanism in $\mathcal{M}^{\tinyi}$ and its symmetric extension to $\mathcal{M}^{\tinyh}$ have the same expected revenue and \vspace{-2mm}
\item[(b)] a symmetric, IC, and IR mechanism in $\mathcal{M}^{\tinyh}$ and its restriction to $\mathcal{M}^{\tinyi}$ have the same expected revenue.\vspace{-1mm}
\end{itemize}
Consequently, optimal mechanisms in models $\mathcal{M}^{\tinyi}$ and $\mathcal{M}^{\tinyh}$ generate the same expected revenue.
\end{itemize}
\end{theorem}

\medskip

In general, the optimization problem for a seller of $n$ heterogeneous objects is quite different from the optimization problem for a seller of $n$ identical units. Theorem~\ref{theo:sym} implies that if the density $f^{\tinyh}$ in $\mathcal{M}^{\tinyh}$ is exchangeable then the seller's revenue-maximization problem is essentially the same as the seller's revenue-maximization problem in $\mathcal{M}^{\tinyi}$ with density $f^{\tinyi} = n! f^{\tinyh}$. Consequently, this equivalence between the two models implies that results in $\mathcal{M}^{\tinyi}$ can be translated into results in $\mathcal{M}^{\tinyh}$, and vice versa. We demonstrate the usefulness of the equivalence in Section~\ref{se:apps}. Without exchangeability, the optimal mechanism in model $\mathcal{M}^{\tinyh}$ need not be symmetric, and hence, Theorem~\ref{theo:sym}(iii) will not hold (but Theorem~\ref{theo:sym}(i) and (ii) continue to hold).

The equivalence between models $\mathcal{M}^{\tinyh}$ and $\mathcal{M}^{\tinyi}$ relies on the decreasing marginal values assumption in the identical-objects model. Indeed, if marginal values are increasing, then $v_i \le v_{i+1}$ for all $v$ whereas, feasibility of a mechanism $(q,t)$ requires that $q_i(v) \ge q_{i+1}(v)$. Thus, a feasible mechanism violates the rank-preserving property under increasing marginal values in the identical-objects model. Hence, Theorem~\ref{theo:sym0} does not hold.

\section{Applications}\label{se:apps}

Our equivalence result on the sale of indivisible objects can be used to transport existing results from the heterogeneous-objects model to the identical-objects model. For instance, approximate expected revenue maximization results for i.i.d. priors in \citet{HN17} imply analogous approximate expected revenue-maximization results in model $\mathcal{M}^{\tinyi}$ for a class of priors. 

We provide another such application of Theorem~\ref{theo:sym} to prove new results for the identical-objects model. Consider an identical-objects model with uniformly-distributed values, i.e.,
\begin{align*}
   f^I(v_1,v_2,\ldots , v_n) \ =\ n!, \qquad\forall \underline{v}+1 \ge v_1\ge v_2\ge \ldots\ge v_n \ge \underline{v} \ge 0
   \end{align*}
Denote this model as $\mathcal{M}^{\tiny{I}}[U]$. Observe that $\mathcal{M}^{\tiny{I}}[U]$ is equivalent to a heterogenous-objects model with i.i.d. values uniformly distributed on $[\underline{v}, \underline{v}+1]$. Then Theorems~5 and 6 of \cite{DDT17} and our Theorem~\ref{theo:sym} imply the following:

\begin{cor}
    \label{cor:u1}
For any integer $n > 0$, there exists a $\underline{v}_0$ such that for all $\underline{v} \ge \underline{v}_0$, the optimal mechanism for selling $n$ units in $\mathcal{M}^{\tiny{I}}[U]$ is a take-it-or-leave-it offer for selling all $n$ units together as a bundle.
\end{cor}

\begin{cor}
    \label{cor:u2}
For any $\underline{v} \ge 0$, there exists an integer $n_0$ such that for all $n \ge n_0$, the optimal mechanism for selling $n$ units in  $\mathcal{M}^{\tiny{I}}[U]$ is {\bf not} a take-it-or-leave-it offer for a bundle of $n$ units.
\end{cor}

Our main interest is in applications of the equivalence result in the other direction.  We provide three such applications where we first establish new results in the identical-objects model, and then extend them to the heterogeneous-objects model using Theorem \ref{theo:sym}. This exploits the fact that the identical-objects model is relatively more tractable than the heterogeneous-objects model, and illustrates that our equivalence result is useful in obtaining new results in the heterogeneous-objects model. 

While the heterogeneous-object model has received much attention in the literature, we believe the identical-object model is also important. One example of a market for identical objects is the sale of financial assets such as U.S. Treasury securities.

\subsection{Object Non-bossiness}\label{se:onb}

The applications require a new condition we call object non-bossiness, which is defined next.\footnote{Object non-bossiness is used in the proof of Theorem \ref{theo:rmon} and is assumed in the hypothesis of Theorem \ref{theo:uic}.}
\begin{defn}
A mechanism $(q,t)$ satisfies {\bf object non-bossiness} if for all $i$, for all $v_{-i}$, and for all
$v_i,v'_i$
\begin{align*}
\Big[q_i(v_i,v_{-i})=q_i(v'_i,v_{-i})\Big]\quad \Longrightarrow\quad  \Big[q_j(v_i,v_{-i})=q_j(v'_i,v_{-i})~\forall~j \in N\Big]
\end{align*}
\end{defn}

In an object non-bossy mechanism, if the allocation probability of the $i^{th}$ unit is the same at two type vectors that differ only in the value of the $i^{th}$ unit, then the allocation probabilities of all units is the same at the two types.\footnote{The idea is similar to {\it agent} non-bossiness introduced by \citet{SS81}.}

Proposition~\ref{pr:nb} in Appendix \ref{sec:papps} shows that for any IC mechanism there exists an object non-bossy and IC mechanism  which is identical to the original mechanism almost everywhere, and generates at least as much revenue everywhere. Thus, an assumption of object non-bossiness is without loss of generality in our environment. 

In some of the applications, we have two specific classes of mechanisms: deterministic and almost deterministic. 
Formally, an allocation rule $q$ is {\bf almost deterministic} if for every $v$ there exists $k \in \{1,\ldots,n\}$
such that $q_i(v) \in \{0,1\}$ for all $i \ne k$. A mechanism $(q,t)$ is (almost) deterministic 
if $q$ is (almost) deterministic.  Proposition~\ref{pr:nb} shows that if an IC mechanism is (almost) deterministic  then its object non-bossy version is also (almost) deterministic.

\subsection{Revenue Monotonicity}\label{se:rmon}

The optimal revenue from the sale of $n$ objects is monotone if the optimal revenue increases when the distribution of the buyer's values increases in the sense of 
first-order stochastic dominance. Monotonicity of the optimal revenue is a desirable 
property as it provides an incentive for the seller to improve her products. It is 
satisfied in the optimal mechanism for the sale of a single object. However, 
as \citet{HR15} show, optimal revenue may not be monotone in the heterogeneous 
objects model. They also show that if the optimal mechanism is symmetric and 
deterministic or if the optimal payment function is submodular, then the optimal 
revenue is monotone in the heterogeneous-objects model.\footnote{In a recent paper, 
\citet{MHN22} show that a restriction to monotone mechanisms can severely reduce 
expected revenue for some classes of distributions.} We provide other sufficient 
conditions that guarantee that the expected revenue from the mechanism 
is monotone in the identical-objects model and the heterogeneous-objects model.

Consider the following definition.
\begin{defn}
A mechanism $(q,t)$ is {\bf revenue monotone} if for every cdf $F$
and every cdf  $\widetilde{F}$, where $\widetilde F$ first-order stochastic dominates $F$, we have
\begin{align*}
\textsc{Rev}(q,t;\widetilde F) \ge \textsc{Rev}(q,t;F)
\end{align*}
\end{defn}
The definition applies to models $\mathcal{M}^{\tinyh}$ and $\mathcal{M}^{\tinyi}$, 
where either $F$ and $\widetilde{F}$ both have support in $\overline{D}^{\tinyh}$ or 
both have support in $\overline{D}^{\tinyi}$.

If a mechanism $(q,t)$ satisfies\footnote{As we assume the existence of densities, 
if (\ref{eq:rtm}) holds for almost all $\hat{v} > v$ then revenue monotonicity is 
satisfied.}
\begin{align}\label{eq:rtm}
t(\hat{v}) \ge t(v)\qquad \forall~\hat{v} > v \
\end{align}
then it satisfies revenue monotonicity, as its expected revenue under a cdf 
$\widetilde F$ is greater than equal to its expected revenue under a first-order 
stochastically-dominated cdf  $F$.\footnote{Note that IC and IR constraints do not 
involve the distribution of values; therefore, if $(q,t)$ is IC and IR under $F$ 
then it is IC and IR under $\widetilde F$.} Thus, if an optimal mechanism satisfies 
(\ref{eq:rtm}) then it is revenue monotone.

Fix an IC mechanism $(q,t)$ and a pair of 
types $v,v'$. Are there sufficient conditions on $q(v)$ and $q(v')$
that imply $t(v) \ge t(v')$? We show that one such condition takes the 
form of majorization. We use this to derive new 
sufficient conditions for revenue monotonicity in both the models.

For any allocation probability vector $q=(q_1,q_2,\ldots,q_n)$, let $q_{[i]}$ be the $i^{\rm th}$ highest element of $q$.  That is, $q_{[1]}\ge q_{[2]}\ge \ldots\ge q_{[n]}$.\footnote{In model $\mathcal{M}^{\tinyi}$, $q_{[i]}=q_i$, for all $i\in N$.} If, for two allocation probability vectors $\hat q,\, q$,
\begin{align*}
\sum_{i=1}^j \hat q_{[i]} &\ge \sum_{i=1}^j  q_{[i]}~\qquad~\forall~j \in \{1,\ldots,n\}
\end{align*}
then $\hat q$ {\bf weakly majorizes} $q$, denoted $\hat q\succ_w  q$.\footnote{If, in addition, $\sum_{i=1}^n q_{[i]}=\sum_{i=1}^n \hat q_{[i]}$ then $\hat q$ {\bf majorizes} $q$. The condition $\sum_{i=1}^n q_{[i]}=\sum_{i=1}^n \hat q_{[i]}$ is not usually satisfied by mechanisms in our setting. } If each of the inequalities above is satisfied with equality, then $\hat q\succ_w  q$ and $ q\succ_w \hat q$; in this case, either $q=\hat q$ or $q$ is a permutation of~$\hat q$.  The $\succ_w$ relation is transitive and incomplete. 

In $\mathcal{M}^{\tinyi}$, since $q_{[i]}=q_i$ for each $i$, a sufficient condition for $\hat{q}\succ_w q$ is that (the cumulative probability distribution function induced by) $\hat{q}$ dominates $q$ by second-order stochastic dominance -- for a formal proof see  Lemma~\ref{le:eqmaj} in Appendix~\ref{sec:onap}.

\begin{prop}\label{pr:schur}
Let $(q,t)$ be an IC mechanism which is either (i) in model $\mathcal{M}^{\tinyi}$ or (ii) in model $\mathcal{M}^{\tinyh}$ and is symmetric. Then, for almost all $v, \hat{v}$, 
\begin{align}\label{eq:schur}
q(\hat{v}) \succ_w q(v) \ \ \Longrightarrow\ \ t(\hat{v}) \ge t(v) 
\end{align}
\end{prop}

\medskip
The intuition behind Proposition~\ref{pr:schur} derives from the IC inequality, \vspace{-2mm}
\begin{align*}
t(\hat{v})-t(v) & \ge v\cdot q(\hat{v})-v\cdot q(v)
\end{align*}
For a mechanism $(q,t)$ in model $\mathcal{M}^{\tinyi}$, we have 
$q_i(v)\ge q_{i+1}(v)$ and $v_i\ge v_{i+1}$. Thus, if $q(\hat{v}) \succ_w q(v)$ 
then the probabilities of acquiring the most valuable bundles are greater at 
$q(\hat{v})$ than at $q(v)$. Hence, the expected value of the allocation under 
$ q(\hat{v})$ is at least as high as the expected value of the allocation under 
$ q(v)$. In consequence, the right-hand expression in the inequality above is 
non-negative and $t(\hat{v})\ge t(v)$.\footnote{\citet{KMS21} study monotone 
functions which majorize or are majorized by a given monotone 
function. They characterize the extreme points of 
such functions and apply their result to several economic problems. 
Our results do not follow from their characterization.}

Consider the following property for an allocation rule.
\begin{defn}
An allocation rule $q$ in $\mathcal{M}^{\tinyi}$, $\mathcal{M}^{\tinyh}$ satisfies {\bf majorization monotonicity} 
if \\ for all $i$, for all $(v_i,v_{-i}), (\hat v_i,v_{-i}) \in D^{\tinyi}, D^{\tinyh}  $, 
\begin{align*}
q_i(\hat v_i,v_{-i}) > q_i(v_i,v_{-i})  \quad \Longrightarrow \quad q(\hat v_i,v_{-i}) \succ_w q(v_i,v_{-i})
\end{align*}
\end{defn}

Note that if $(q,t)$ is IC, then 
$q_i(\hat{v}_i,v_{-i}) > q_i(v_i,v_{-i})$ implies $\hat{v}_i > v_i$.
Hence, majorization monotonicity (for an IC mechanism) is weaker than 
requring $\hat{v}_i > v_i$ implies $q(\hat v_i,v_{-i}) \succ_w q(v_i,v_{-i})$.
Theorem~\ref{theo:rmon}(a) below establishes that 
majorization monotonicity is sufficient for revenue monotonicity. 
It also provides a sufficient condition for majorization monotonicity (and, hence, for
revenue monotonicity). 

\begin{theorem} \label{theo:rmon} 
Suppose that $(q,t)$ is an IC mechanism in model $\mathcal{M}^{\tinyi}$ or a 
symmetric IC mechanism in model $\mathcal{M}^{\tinyh}$.
$\qquad$\\\vspace{-10mm}
\begin{itemize}
\item[(a)] If $q$ satisfies majorization monotonicity then $(q,t)$ is revenue monotone. \vspace{-3mm}
\item[(b)] If $q$ is almost deterministic, then it satisfies majorization monotonicity, and 
hence, $(q,t)$ is revenue monotone.
\end{itemize}
\end{theorem}

\begin{remark}\label{re:revmon} \em
By Proposition \ref{prop:sym2}, if the distribution is exchangeable
then there exists a symmetric mechanism that is optimal in the heterogeneous-objects 
model. Hence, such an optimal mechanism is revenue monotone if it is either 
(a) majorization monotone or (b) almost deterministic. In other words, if $F^{\tinyh}$ is an exchangeable distribution, then for every $\widetilde{F}^{\tinyh}$ that first-order stochastic dominates $F^{\tinyh}$, 
the optimal revenue under $\widetilde{F}^{\tinyh}$ is no less than the optimal revenue under $F^{\tinyh}$. Note
that $\widetilde{F}^{\tinyh}$ need not be an exchangeable distribution. \hfill $\Box{}$
\end{remark}

Theorem \ref{theo:rmon}(b) strengthens the result in \citet{HR15}, who showed that the 
optimal mechanism in model $\mathcal{M}^{\tinyh}$ is revenue monotone if it is symmetric and 
deterministic. It is difficult to find sufficient conditions on the primitives of the model that guarantee
existence of an optimal mechanism which is symmetric and deterministic.  However, as shown next, when $n=2$ there is a simple condition on the distribution of values that guarantees that the optimal 
mechanism is almost deterministic. 

Consider the following condition on the density of buyer types, which was introduced by \citet{MM88}:
\begin{align}\label{eq:sch}
3f^{\tinym}(v)+v\cdot \nabla f^{\tinym}(v)\ge 0\qquad \forall~v\in \overline{D}^{\tinym},\ M=H \mbox{ or } I
\end{align}
The uniform family of distributions, the truncated exponential distribution, and a family of Beta distributions
satisfy condition (\ref{eq:sch}). When there are two objects and $\underline{v}=0$, condition (\ref{eq:sch}) is sufficient for the existence of an optimal mechanism which is almost deterministic. This is shown in Proposition 1 of \citet{Pa11b} for model $\mathcal{M}^{\tinyh}$ and in Proposition 1 of \citet{BM22} for model $\mathcal{M}^{\tinyi}$.  Thus, we have the following result.

\begin{cor}\label{cor:rmon}
  Suppose that $n=2$, $\underline{v}=0$, (\ref{eq:sch}) is satisfied, and $f^{\tinym}$ is continuously
  differentiable and positive for $M=H$ or $I$. Then\vspace{-3mm}
  \begin{itemize}
\item[(a)] An optimal mechanism in model $\mathcal{M}^{\tinyi}$ is revenue monotone. \vspace{-3mm}
\item[(b)]
Further, if $f^{\tinyh}$ is exchangeable, then an optimal mechanism in model $\mathcal{M}^{\tinyh}$ is revenue monotone.
\end{itemize}
  \end{cor}

Note that under the hypothesis of Corollary~\ref{cor:rmon}, revenue increases with any distribution $f'$ that dominates $f^{\tinym}$ by first-order stochastic dominance, whether or not $f'$ satisfies (\ref{eq:sch}).

Whether, for the case $n\ge3$, there is a simple condition on the distribution of values that ensures the existence of an optimal mechanism that satisfies majorization monotonicity is an open question.

\subsection{Pricing Mechanisms}\label{se:price}

In this subsection and the next, we restrict attention to deterministic mechanisms.  As shown by \cite{T04} and  \cite{MV06}, this is not without loss of generality as an optimal mechanism might be stochastic. Conditions on the primitives of the model that guarantee that there an optimal mechanism which is deterministic are not known.\footnote{However, see \cite{BM22} for conditions that ensure that a deterministic mechanism is optimal when $n=2$.} At the same time, stochastic mechanisms are rarely used in practice. One explanation is given by \citet{LM02}, who argue that commitment (an implicit assumption in mechanism design) to stochastic mechanisms is more difficult to achieve and verify than commitment to deterministic mechanisms. This is especially true in one-shot interactions. Hence, even though theoretically restrictive, deterministic mechanisms can be of practical interest.

Throughout, we refer to an optimal mechanism in the set of all deterministic mechanisms as an {\sl optimal deterministic mechanism}. We refer to an optimal mechanism in the set of all symmetric, deterministic mechanisms as an {\sl optimal symmetric deterministic mechanism}.

Theorem~\ref{theo:orderedi} below provides a sufficient condition on the probability distribution of buyer values under which optimal prices in the identical-objects model cannot be increasing. This in turn implies that the optimal prices in the heterogeneous-objects model with i.i.d. buyer values cannot be supermodular (Theorem~\ref{theo:orderedm}).

A first step in the proof is to show that any deterministic mechanism can be expressed as a  pricing
mechanism, which is defined next.\vspace{-2mm}

\begin{defn}\label{de:pm}
A deterministic mechanism $(q,t)$ for the identical-objects model defined on the type space 
$D^{\tinyi}$ is a {\bf pricing mechanism} if there exists prices $p_0=0, p_1,\ldots,p_n  \in [0,\overline{v}]$, such that for all $v \in D^{\tinyi}$,\vspace{-2mm}
\begin{alignat*}{1}
	t(v) = \sum_{k=0}^{k(v)}p_{k}
\end{alignat*}
where $k(v):= \sum_{i=1}^n q_i(v)$.
\end{defn}\vspace{-2mm}

Suppose that $(q,t)$ is a deterministic, IC and IR mechanism with {\bf full range,} i.e., for each $k=0,1,\ldots,n$ there exists a type $v^k$ that is allocated $k$ units. Then the pricing mechanism implementation of $(q,t)$ has prices $p_k=t(v^k)-t(v^{k-1})$. That pricing mechanism implementations exist for deterministic mechanisms without full range is established next.

\begin{prop}\label{prop:price}
If $(q,t)$ is a deterministic, IC, and IR mechanism in model $\mathcal{M}^{\tinyi}$, then it is a pricing mechanism 
with prices $p_0=0,p_1,\ldots,p_n \in [0,\overline{v}]$ such that 
\begin{align}
	u(v) &\ge \sum_{i=0}^k v_i - \sum_{i=0}^k p_i ~\qquad~\forall~v \in D^{\tinyi}~\forall~k \in \{0,1,\ldots,n\}\label{eq:icp}
\end{align}
where $v_0:=0$.
\end{prop}

Proposition~\ref{prop:price} implies that even if an IC and IR mechanism does not have full range, it can be implemented by defining prices for {\sl every} unit and letting the buyer choose the payoff maximizing units. \citet{HR15} prove a similar result for the heterogeneous-objects model with an unbounded type space.
Due to unbounded type space, prices of objects in their model can be infinite. In contrast, the prices in Definition~\ref{de:pm} are bounded above by $\overline{v}$. The proof of Proposition~\ref{prop:price} is in  Appendix \ref{sec:onap}.

\begin{figure}
	\centering
	\includegraphics[width=3in]{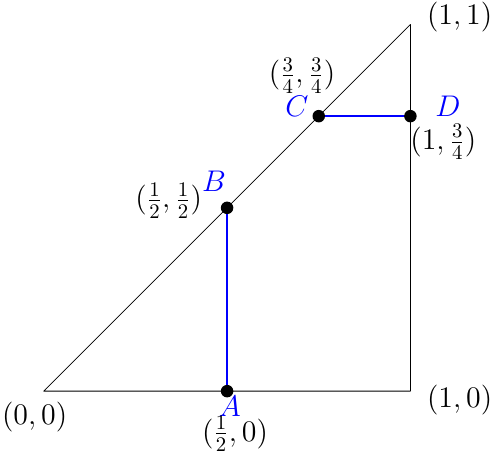}
	\caption{Increasing prices}
	\label{fig:inc_pr}
\end{figure}
Since marginal values are decreasing, it is tempting to think that prices are decreasing in an optimal pricing mechanism. As shown next, this intuition is incorrect.

\begin{example}
\label{ex:incpric}
{\em Consider an example with two units. The type space consists of the blue line segments $AB$ and $CD$ in Figure \ref{fig:inc_pr}. The values of the buyer are on either $AB$ with probability $\frac{1}{2}$ or $CD$ with probability $\frac{1}{2}$. Further, conditional on values belonging to $AB$ or $CD$, the values are distributed uniformly on each line segment. It may be verified that the optimal prices are $p_1=\frac{1}{2}$ and $p_2=\frac{3}{4}$; types on $AB$ buy one unit and types on $CD$ buy both the units, resulting in an expected revenue of $\frac{7}{8}$.\footnote{The other two candidates for optimal prices, $(\frac 12, \frac 12)$ and $(\frac 34, \frac 34)$, each yield an expected revenue of $\frac 3 4 $.}  \hfill$\blacksquare$ }
\end{example}

Thus, optimal mechanisms need not have decreasing prices. Next, we give a simple sufficient condition under which the optimal mechanism does not have increasing prices (which ensures decreasing prices for two units). 

For probability distribution $F$ over buyer values in the identical-objects model, let $F_k$ denote the marginal distribution of the value for the $k$-th unit. If $\frac{1-F_k(t)}{1-F_{k+1}(t)}$ is increasing in $t$ then $F_k$ {\bf hazard-rate dominates} $F_{k+1}$. Theorem~\ref{theo:orderedi} shows that under this condition prices cannot be increasing in any optimal pricing mechanism. The proof does not require the existence of a probability density function over values.

\begin{theorem}\label{theo:orderedi}
Suppose $F$ is a probability distribution over buyer values in model $\mathcal M^{\tinyi}$ such that $F_k$ hazard-rate dominates $F_{k+1}$ for all $k=1,\ldots,n-1$. Then there is no optimal deterministic mechanism with respect to $F$ with prices $p_0,p_1,\ldots,p_n$ such that \vspace{-2mm}
\begin{alignat*}{1}
	p_1 \le p_2 \le \ldots \le p_n
\end{alignat*}\vspace{-2mm}
with at least one strict inequality.
\end{theorem}

Suppose that a buyer may adopt multiple identities without detection by the seller. Then prices $p_1 \le \ldots \le p_k<p_{k+1} \le\ldots \le p_n$  cannot be implemented as a buyer can buy all $n$ units at a price strictly less than $\sum_{i=1}^n p_i$ by purchasing $k$ and $n-k$ units separately. In other words, increasing prices are not {\sl false-name proof} (see \cite{YSM04}). Under the hypothesis of Theorem~\ref{theo:orderedi}, this requirement imposed by false-name proofness is without cost to the seller.  

Let $(q^{\tinyh},t^{\tinyh})$ be a deterministic, symmetric, IC and IR mechanism in model $\mathcal{M}^{\tinyh}$. By Theorem \ref{theo:sym}, its restriction to $D^{\tinyi}$ defines an IC and IR mechanism $(q^{\tinyi},t^{\tinyi})$ in model $\mathcal{M}^{\tinyi}$. By Proposition \ref{prop:price}, $(q^{\tinyi},t^{\tinyi})$ is a pricing mechanism with prices $p_0=0,p_1,\ldots,p_n$. Let $v \in D^{\tinyh}$ be a type such that $q^{\tinyh}(v)$ allocates bundle $S$ in $\mathcal{M}^{\tinyh}$. There exists a permutation $\sigma$ such that $v^{\sigma} \in D^{\tinyi}$. By symmetry, $t^{\tinyh}(v)=t^{\tinyh}(v^{\sigma})$ and $q^{\tinyh}(v^\sigma)$ allocates the permuted bundle $S^{\sigma}$. But $q^{\tinyh}(v^\sigma)=q^{\tinyi}(v^\sigma)$ and since $(q^{\tinyi},t^{\tinyi})$ is a pricing mechanism, we have 
\begin{align*}
	t^{\tinyh}(v^{\sigma}) = t^{\tinyi}(v^{\sigma}) = \sum_{j=1}^{|S^{\sigma}|}p_j = \sum_{j=1}^{|S|}p_j = t^{\tinyh}(v)
\end{align*}
Hence, the prices of a symmetric, deterministic, IC and IR mechanism in model $\mathcal{M}^{\tinyh}$
can be described by prices $p_0=0,p_1,\ldots,p_n$. For notational convenience, let $P(S) \equiv \sum_{j=1}^{|S|}p_j$ be the price of bundle $S$.  

\begin{defn}\label{de:supmod} 
	A deterministic mechanism $(q^{\tinyh},t^{\tinyh})$ with prices $\{P(S)\}_S$ in model $\mathcal M^{\tinyh}$  is {\bf supermodular} if 
	\begin{align*}
		P(S\cup\{k\})-P(S) & \le P(T\cup \{k\})- P( T), \qquad \forall S\subsetneq T\subseteq N, \ \forall k\notin T
	\end{align*}
	A mechanism is strictly supermodular if at least one of the above inequalities is strict.
	\end{defn}

Supermodularity is well defined for deterministic, symmetric mechanisms as $P(S):=\sum_{j \in S}p_j$ is defined for every bundle $S\subseteq N$, even if no type is allocated $S$. Note that supermodularity is equivalent to  $p_0=0 \le p_1 \le \ldots \le p_n$ and strict supermodularity is equivalent to at least requiring one of these inequalities to be strict. 

An implication of Theorem~\ref{theo:orderedi} is that  if values of objects are distributed independently and identically in the heterogeneous model, then the prices in the optimal mechanism cannot be supermodular.

\begin{theorem}\label{theo:orderedm}
	Consider a heterogeneous-objects model $\mathcal M^{\tinyh}$ in which the values of the objects are distributed i.i.d. Then there is no optimal symmetric deterministic mechanism which is strictly supermodular.
\end{theorem}	

\cite{BNA18} showed that even if $F$ is i.i.d.\,in model $\mathcal{M}^{\tinyh}$, the optimal deterministic mechanism need not be symmetric when there are three or more objects for sale.\footnote{Babaioff et al.'s result is about optimal deterministic mechanisms. By Proposition~\ref{prop:sym2}, we know that there exists an optimal mechanism that is symmetric but it need not be stochastic.} Identifying conditions on i.i.d. $F$ under which the optimal deterministic mechanism is symmetric is an open question.

For the case $n=2$, Theorem~\ref{theo:orderedm} implies that optimal deterministic prices are submodular.\footnote{From \cite{BNA18}  we know that the optimal deterministic mechanism is symmetric when $n=2$.} As \cite{MV06} point out, if buyers can buy or sell objects freely, then only submodular prices can be implemented. Moreover, submodular prices are also false-name proof. 

\subsection{Upper-set Incentive Compatibility}\label{se:dic}

Next, we show that a subset of IC constraints imply full incentive compatibility for deterministic mechanisms. We establish that upper-set incentive constraints (i.e., incentive constraints between $v$ and $v'$ whenever $v\ge v'$ or $v'\ge v$) are sufficient to imply all IC constraints in deterministic and object non-bossy mechanisms in the identical-objects model. Theorem~\ref{theo:sym0} implies that this relaxation of the set of IC constraints applies to symmetric, rank-preserving and deterministic mechanisms in the heterogeneous-objects model whether or not the distribution of types is exchangeable.\footnote{These results extend to many buyers provided the concept of incentive compatibility is dominant strategy.}

Theorem~\ref{theo:uic} below shows that upper-set IC constraints are sufficient for IC for deterministic mechanisms in model $\mathcal{M}^{\tinyi}$. It is proved for a broad class of type spaces which include $\overline{D}^{\tinyi}$ and~$D^{\tinyi}$. 

Let $D$ be an arbitrary type space in the identical-objects model.
Denote the IC constraint that type $v$ does not gain by misreporting type $v'$ as $v \rightarrow v'$. Denote $v \rightarrow v'$ and $v' \rightarrow v$ by $v \leftrightarrow v'$. For every $v \in D$, the {\bf upper set} of $v$ is defined as
\begin{align*}
T(v) &= \{v' \in D: v'_i \ge v_i~\forall~i \in N\}
\end{align*}

\begin{defn}
A mechanism $(q,t)$ for an identical-objects model defined on $D$ is {\bf upper-set incentive compatible (UIC)} if for every $v \in D$ and every $v' \in T(v)$, the IC constraints $v \leftrightarrow v'$ hold.
\end{defn}

The sufficiency of UIC for IC for deterministic mechanisms in the identical-objects model is shown next.

\begin{theorem}\label{theo:uic} 
Every deterministic, object non-bossy, and upper-set incentive compatible
mechanism in an identical-objects model defined on $\overline{D}^{\tinyi}$ is incentive compatible.
\vspace{-2mm}
\end{theorem}

This theorem does not apply to stochastic mechanisms or to bossy deterministic mechanisms, as shown in Example~\ref{ex:uic} in Appendix~\ref{sec:onap}.

\vspace{-2mm}

\begin{remark}\label{rem:simpleric}\em
The multidimensional mechanism-design problem has proved analytically intractable in part because IC constraints may bind in all directions. In contrast, only local downward IC constraints bind in the one-dimensional problem. Theorem~\ref{theo:uic} identifies a multidimensional setting where only upward and downward IC constraints bind. Further, as noted in Remark~\ref{rem:localic} below, UIC can be weakened to local upward and downward IC constraints.

Theorem \ref{theo:uic} is proved for more general types spaces than $D^{\tinyi}$, including finite type spaces. In finite type spaces, an optimal deterministic mechanism is a solution to a finite-dimensional integer program. This may be useful in obtaining results on multidimensional screening problems. One such potential application is the primal-dual formulation of a multiproduct monopolist in \cite{CDRT24}. While they consider general mechanisms, Theorem~\ref{theo:uic}  implies that a restriction to deterministic mechanisms would greatly simplify the tree structure of the binding IC constraints (see Figure 1 in \cite{CDRT24}). The resulting  reduction in the number of constraints in the primal and the number of variables in the dual may provide further insight into optimal deterministic mechanisms. 
\hfill $\Box{}$\vspace{-5mm}
\end{remark}

\begin{remark}\label{rem:localic} \em We contrast the sufficiency of UIC for IC from 
previous work that has shown the sufficiency of a reduced set of incentive constraints in multidimensional environments. A mechanism satisfies {\em local IC constraints} 
if for every $v$, there exists an $\epsilon > 0$ such that for all 
$v'$ in an $\epsilon$-ball around $v$, the IC constraints 
$v \rightarrow v'$ and $v' \rightarrow v$ hold. \citet{Ca12} shows 
that local IC constraints imply all IC constraints in convex type spaces; this 
holds for stochastic mechanisms also.
This result has been extended to non-convex type spaces by \citet{MPR16} and \citet{KR21}. 
UIC does not correspond to any notion of locality, and 
hence, the set of redundant IC constraints identified by UIC is quite different. In fact, we can use \citet{Ca12} to strengthen Theorem~\ref{theo:uic} by weakening the requirement of UIC to local UIC.\footnote{If $D$ is convex, the upper set of any type $v$ is a convex set. Hence, satisfying all IC constraints in the upper set of $v$ is equivalent to satisfying all local IC constraints in the upper set of $v$. Similarly, if $v \ge v' \ge v''$, and IC constraints $v \leftrightarrow v', v' \leftrightarrow v''$ hold, then a straightforward argument shows that the IC constraint $v \leftrightarrow v''$ holds. Thus, it is possible to define a (weaker) local notion of upper set IC which implies~IC.} \hfill $\Box{}$
\end{remark}

\begin{remark}\label{rem:relax} \em
 In mechanism-design problems, whenever a subset of IC constraints is dropped, the set of mechanisms satisfying the smaller set of IC constraints is usually larger. This is called a {\em relaxed program}. The aim is to find conditions under which an optimal solution of the relaxed program is also an optimal solution of the original problem (see, for instance, \citet{Ar00}). 

In contrast, we show that UIC constraints imply all IC constraints for any deterministic and object non-bossy mechanism. Thus, for any optimization program in this setting (be it revenue maximization or some other objective function), it is without loss of generality to consider a strictly smaller set of IC constraints, the set of UIC constraints. \hfill $\Box{}$
\end{remark}

Next, we derive an analog of Theorem \ref{theo:uic} for model $\mathcal{M}^{\tinyh}$. As we consider only symmetric mechanisms, it is sufficient to define UIC on $D(\sigma^{\tinyi})$.\vspace{-2mm}

\begin{defn}
A symmetric and rank-preserving mechanism $(q,t)$ for the heterogeneous-objects model defined on type space $D^{\tinyh}$ is {\bf upper-set incentive compatible (UIC)} if for every $v \in D(\sigma^{\tinyi})$ and
for every $v' \in T(v) \cap D(\sigma^{\tinyi})$, the 
IC constraints $v \leftrightarrow v'$ hold.\vspace{-2mm}
\end{defn}

A symmetric mechanism $(q,t)$ in the heterogeneous-objects model is {\bf object non-bossy} if $q_i(v_i,v_{-i})=q_i(v'_i,v_{-i})$ implies $q(v_i,v_{-i})=q(v'_i,v_{-i})$ for every $(v_i,v_{-i}),(v'_i,v_{-i}) \in D(\sigma^{\tinyi})$. 

Theorem \ref{theo:uic} and Theorem \ref{theo:sym0} imply the following:\vspace{-2mm}
\begin{theorem} 
\label{theo:suic}
Every symmetric, deterministic, object non-bossy, rank-preserving and upper-set incentive compatible
mechanism in a heterogeneous-objects model defined on $D^{\tinyh}$ is incentive compatible.
\end{theorem}

Since the definition of symmetry only applies to strict types, Theorem~\ref{theo:suic} 
does not extend to $\overline{D}^{\tinyh}$. However, if $(q,t)$ is a 
symmetric, rank-preserving, deterministic, object non-bossy,  UIC 
mechanism defined on $\overline{D}^{\tinyh}$, then its restriction to $D^{\tinyh}$ is 
IC by Theorem~\ref{theo:suic}. Using Lemma~\ref{le:int}, there exists another 
IC mechanism $(q',t')$ defined on $\overline{D}^{\tinyh}$ that coincides with $(q,t)$
almost everywhere, and hence, generates the same expected revenue as $(q,t)$.

In the heterogeneous-objects model, IC and symmetric mechanisms are rank preserving (by Theorem~\ref{theo:sym0}). However, UIC and symmetric mechanisms need not be rank preserving, and hence we need to assume rank preserving in Theorem~\ref{theo:suic}. Note that the assumption of exchangeability is not required for Theorem~\ref{theo:suic} as incentive compatibility does not depend on the probability distribution of values.

\section{Discussion}\label{sec:disc}

We have provided new results in the heterogeneous-objects model by first obtaining new results in the identical-objects model and then using our equivalence result. One can also translate known results in the identical-objects setting to the heterogeneous-objects setting. In a model with two identical objects, \citet{BM22} obtain a sufficient condition for the existence of an optimal mechanism that is deterministic. This result can be directly applied to selling two heterogeneous objects with an exchangeable distribution of values.

While our presentation is in terms of selling indivisible objects, the results apply to other 
settings. For example, consider the following two models: \vspace{-2mm}

\begin{enumerate}

\item {\sc Model I.} A seller offers one durable product which, depending on its quality level, may be consumed for up to 
$n$ periods. An object with quality level $i$ lasts $i$ periods. The seller sells (at most) one object to a buyer at 
the beginning of the first period at one of the $n$ quality levels; no sales take place at any later time period. 
If a buyer purchases an object of quality $i$ at the beginning of the first period, then she consumes it in each of the periods $1,2,\ldots,i$. 
The value of consuming the product in the $n$ periods is $(v_1,\ldots,v_n)$. Owing to discounting, which need not 
be same across periods, $v_1 \ge v_2 \ge \ldots \ge v_n$. So, $v_i$ is the marginal value of increasing the quality 
level from $i-1$ to $i$. Note that the product can be consumed in period $i$ only if it is consumed in period $i-1$ 
and hence $q_{i-1}(\cdot)\ge q_i(\cdot)$, where $q_i(\cdot)$ is the probability of consuming
the product in period $i$.

\item {\sc Model H.} This is a one-period model in which a seller offers $n$ different products, each of which lasts 
one period. So, $v_i$ denotes the value for product $i$ to the buyer. The buyer has additive values over any subset 
of the products.
\end{enumerate}

\vspace{-3mm}
As long as the products in {\sc Model H} are {\sl ex-ante symmetric}, our results imply that
the two models are equivalent.

\newpage

\begin{appendices}
\section{Omitted Proofs}

\subsection{Proofs of Section \ref{se:srpm}}

\noindent
{\sc Proof of Lemma~\ref{le:int}:} Let $(\bar q(v), \bar t(v)) \equiv ( q(v),  t(v)), \ \forall v\in D^{\tinym}$. For any $v \in \overline{D}^{\tinym}  \setminus D^{\tinym}$, take a sequence $\{v^k\}_k$ in $D^{\tinym}$ that converges to $v$. (As $D^{\tinym}$ is dense in $\overline{D}^{\tinym}$, for each
$v \in \overline{D}^{\tinym} \setminus D^{\tinym}$, there exists such a sequence.)  As already noted, for each $v$, $t(v)$ is bounded above and below and $q_i(v^k) \in [0,1]$. Thus, $\{q(v^k),t(v^k)\}_k$ is a bounded sequence and hence, it has an accumulation point. Set $(\bar q(v), \bar t(v))$ equal to an accumulation point of this sequence. For every $v \in \overline{D}^{\tinym} \setminus D^{\tinym}$,  $(\bar q(v),\bar t(v))$ is an accumulation point of outcomes of a sequence of types in $D^{\tinym}$. Therefore, as $(q,t)$ is IC and IR on $D^{\tinym}$,
and the buyer's payoff function is continuous in $v$, it follows that $(\bar q, \bar t)$ is IC and IR on $\overline{D}^{\tinym}$.
\hfill$\blacksquare$

\medskip
\noindent
{\sc Proof of Theorem~\ref{theo:sym0}:} Let  $(q,t)$ be a  symmetric mechanism in $\mathcal{M}^{\tinyh}$. \\
\underline{$(i) \Rightarrow (ii)$:}
Assume that $(q,t)$ is IC on $D^{\tinyh}$. Fix $i$ and $j$.
Let $v\in D^{\tinyh} $ and let $\sigma$ be the permutation such that $\sigma(i)=j,\sigma(j)=i$ and $\sigma(k)=k$
for all $k \notin \{i,j\}$. We have
\begin{align*}
0 &= t(v) - t(v^{\sigma})~\,\qquad\qquad\qquad\qquad\qquad~\textrm{(by symmetry of $(q,t)$)} \\
&\le v \cdot (q(v) - q(v^{\sigma})) ~\qquad\qquad\qquad\qquad~\textrm{(by IC of $(q,t)$)}\\
&= v_i (q_i(v)-q_i(v^{\sigma})) + v_j (q_j(v) - q_j(v^{\sigma})) \\
&= (v_i - v_j) (q_i(v) - q_j(v))
\end{align*}
where the the last two equalities follow from symmetry. Thus, if $v_i > v_j$, then $q_i(v) \ge q_j(v)$. Hence, $(q,t)$ is rank-preserving.

\noindent \underline{$(ii) \Rightarrow (i)$:} 
Pick any $v \in D(\sigma)$ and $\hat{v} \in D(\hat \sigma)$. These map to $v^{\sigma}, \hat{v}^{\hat{\sigma}} \in D(\sigma^{\tinyi})$ such that for every $i$,\vspace{-3mm}
\begin{align}
v^{\sigma}_i &= v_{\sigma(i)}, ~~~~\hat{v}^{\hat{\sigma}}_i = \hat{v}_{\hat{\sigma}(i)} \label{eq:s1}
\end{align}

\vspace{-2mm}
We know that
\begin{align}
\sum_{i=1}^n v_i q_i(v) - t(v) &= \sum_{i=1}^n v_{\sigma(i)} q_{\sigma(i)}(v) - t(v) \nonumber \\
&= \sum_{i=1}^n v^{\sigma}_i q_i(v^{\sigma}) - t(v^{\sigma})~\qquad~\textrm{(by symmetry of $(q,t)$ and (\ref{eq:s1}))} \nonumber \\
&\ge \sum_{i=1}^n v^{\sigma}_i q_i(\hat{v}^{\hat{\sigma}}) - t(\hat{v}^{\hat{\sigma}}) \nonumber\\
&= \sum_{i=1}^n v^{\sigma}_i q_{\hat\sigma(i)}(\hat{v}) - t(\hat{v})  \label{eq:s3},~\qquad~\textrm{(by symmetry of $(q,t)$)}
\end{align}
where the inequality follows as  $v^{\sigma},\hat{v}^{\hat{\sigma}} \in D(\sigma^{\tinyi})$ and $(q,t)$ restricted to $D(\sigma^{\tinyi})$ is IC.

Note that
\begin{align*}
& v^{\sigma}_1 > v^{\sigma}_2 > \ldots > v^{\sigma}_n \ \quad\qquad\qquad\qquad\quad~\textrm{(since $v^{\sigma} \in D(\sigma^{\tinyi})$)} \\
& q_{\hat\sigma(1)}(\hat{v}) \ge q_{\hat\sigma(2)}(\hat{v}) \ge \ldots \ge q_{\hat\sigma(n)}(\hat{v}) \qquad~\textrm{(since $(q,t)$ is rank-preserving, $\hat{v} \in D(\hat\sigma)$, and (\ref{eq:sigma}))}
\end{align*}
As $(v_{\hat\sigma(1)},v_{\hat\sigma(2)},\ldots,v_{\hat\sigma(n)})$ is a permutation of  $(v^{\sigma}_1, v^{\sigma}_2, \ldots, v^{\sigma}_n)$, these inequalities imply that\footnote{See also the rearrangement inequality in Theorem 368 of \citet{HLP52}.}
\begin{align}
\sum_{i=1}^n  v^{\sigma}_i q_{\hat\sigma(i)}(\hat{v}) &\ge \sum_{i=1}^n v_{\hat{\sigma}(i)} q_{\hat\sigma(i)}(\hat{v}) \label{eq:s4}
\end{align}
Using (\ref{eq:s3}) and (\ref{eq:s4}), we have
\begin{align*}
\sum_{i=1}^n v_i q_i(v) - t(v) &\ge  \sum_{i=1}^n v_{\hat{\sigma}(i)} q_{\hat\sigma(i)}(\hat{v})  - t(\hat{v}) \\
&= \sum_{i=1}^n v_i q_i(\hat{v}) - t(\hat{v}),
\end{align*}
which is the desired IC constraint.
\hfill$\blacksquare$

\medskip
\noindent
{\sc Proof of Lemma~\ref{le:symex}:} Relabelling the objects if necessary, 
assume that $(q,t)$ is defined on $D(\sigma^{\tinyi})$. Let $(q^s,t^s)$ be the 
symmetric extension of $(q,t)$. As $(q,t)$ is rank preserving on 
$D(\sigma^{\tinyi})$, $(q^s,t^s)$ is rank preserving on $D^{\tinyh}$.  
As $(q^s,t^s)=(q,t)$ on $D(\sigma^{\tinyi})$ and $(q,t)$ is IC on 
$D(\sigma^{\tinyi})$, we conclude that $(q^s,t^s)$ is IC on $D^{\tinyh}$ 
(by Theorem \ref{theo:sym0}).

For any $v\in D(\sigma^{\tinyi})$, $(q^s(v),t^s(v))=(q(v),t(v))$. Thus, 
$(q^s,t^s)$ is IR on $D(\sigma^{\tinyi})$. That $(q^s,t^s)$ is IR follows from 
the fact that the payoff of any type $v\in  D(\sigma)$ is the same as the 
payoff of type $v^\sigma\in D(\sigma^{\tinyi})$. \hfill$\blacksquare$

\medskip
\noindent
{\sc Proof of Theorem~\ref{theo:sym}:} \\
(i) \& (ii) Note that any mechanism in model $\mathcal{M}^{\tinyi}$ is rank preserving due to feasibility restriction~(\ref{eq:idrp}). Thus, by Lemma \ref{le:symex} the symmetric extension of an IC and IR mechanism in model $\mathcal{M}^{\tinyi}$ is a rank-preserving, IC and IR mechanism on $D^{\tinyh}$; this symmetric mechanism can be extended to $\overline{D}^{\tinyh}$, i.e., to model~$\mathcal{M}^{\tinyh}$, by Lemma~\ref{le:int}. Conversely, if $(q,t)$ is a symmetric, IC, and IR mechanism in model $\mathcal{M}^{\tinyh}$, then it is rank preserving by Theorem~\ref{theo:sym0}. Hence, the restriction of $(q,t)$ to $D(\sigma^{\tinyi})$ defines an IC and IR mechanism for model $\mathcal{M}^{\tinyi}$ (since rank preserving implies that the feasibility restriction (\ref{eq:idrp}) holds). 

(iii) As $f^{\tinyh}$ is exchangeable, Proposition \ref{prop:sym2} implies that there exists a symmetric and rank-preserving  mechanism that is optimal in model $\mathcal{M}^{\tinyh}$.  Let $(q^{\tinyh},t^{\tinyh})$ be this optimal mechanism in model $\mathcal{M}^{\tinyh}$ and let $(q^{\tinyi},t^{\tinyi})$ be the corresponding IC and IR mechanism for model $\mathcal{M}^{\tinyi}$ obtained by restricting $(q^{\tinyh},t^{\tinyh})$ to $D(\sigma^{\tinyi})$. As $(q^{\tinyh},t^{\tinyh})$ is symmetric, we have
\begin{align}
\textsc{Rev}(q^{\tinyh},t^{\tinyh};f^{\tinyh}) &= 
n! \int_{D^{\tinyi}}t^{\tinyh}(v) f^{\tinyh}(v) dv 
= \int_{D^{\tinyi}}t^{\tinyi}(v) f^{\tinyi}(v) dv 
= \textsc{Rev}(q^{\tinyi},t^{\tinyi};f^{\tinyi}) \label{eq:revdmv}
\end{align}

As $(q^{\tinyh},t^{\tinyh})$ is optimal in $\mathcal{M}^{\tinyh}$, $(q^{\tinyi},t^{\tinyi})$ must be optimal in $\mathcal{M}^{\tinyi}$. To see this, suppose that some other mechanism $(q^{\tinyi'},t^{\tinyi'})$ yields a strictly higher revenue than $(q^{\tinyi},t^{\tinyi})$ in $\mathcal{M}^{\tinyi}$. Let $(q^{\tinyh'},t^{\tinyh'})$ be the symmetric extension of  $(q^{\tinyi'},t^{\tinyi'})$ to $\mathcal{M}^{\tinyh}$. Then we have 
\begin{align*}
\textsc{Rev}(q^{\tinyh},t^{\tinyh};f^{\tinyh})	& =  \textsc{Rev}(q^{\tinyi},t^{\tinyi};f^{\tinyi}) < \textsc{Rev}(q^{\tinyi'},t^{\tinyi'};f^{\tinyi})  = \textsc{Rev}(q^{\tinyh'},t^{\tinyh'};f^{\tinyh})
\end{align*}
where the equalities follow from (\ref{eq:revdmv}). But this contradicts the assumption that  $(q^{\tinyh},t^{\tinyh})$ is optimal in $\mathcal{M}^{\tinyh}$. Thus, optimal mechanisms in the two models yield the same expected revenue.  \hfill$\blacksquare$

\subsection{Proofs of Section \ref{se:apps}}\label{sec:papps}

\medskip\noindent

We prove that object non-bossiness is without loss of generality. We do this for an arbitrary type space $D^*$ with decreasing marginal values, which can be finite or infinite. In particular, $D^*$ can be $D^{\tinyi}$ or $\overline{D}^{\tinyi}$.

\begin{prop}\label{pr:nb}
Suppose $(q,t)$ is an IC and IR mechanism in the identical-objects model defined on an arbitrary type space $D^*$ with decreasing marginal values. Then, there exists an IC, IR, and object non-bossy mechanism $(q^\sharp,t^\sharp)$ such that:\vspace {-2mm}
\begin{enumerate}
\item[(i)]$t^\sharp(v) \ge t(v)$ and $u^\sharp(v)=u(v)$ for all $v \in D^*$.\vspace{-2mm}
\item[(ii)] If $D^*$ is convex, then $(q^\sharp(v),t^\sharp(v)) = (q(v),t(v))$ for almost all $v \in D^*$.\vspace {-2mm}
\item[(iii)] If $(q,t)$ is deterministic (almost deterministic), then $(q^\sharp,t^\sharp)$ can be 
chosen to be deterministic (almost deterministic).
\end{enumerate}
\end{prop}
\noindent {\sc Proof.} We provide a proof of parts (i) and (ii) for an arbitrary IC mechanism, and indicate the changes needed for the proof to work for deterministic and almost deterministic mechanisms, i.e., for part (iii). 

\noindent
{(i): }
Let $(q,t)$ be an IC mechanism defined on $D^*$ with corresponding utility function $u$. 
An $x \in [0,1]^n$ with $x_1 \ge \ldots \ge x_n$ is a {\bf subgradient}\footnote{For the proof of 
deterministic and almost deterministic mechanisms, only consider subgradients which are 
deterministic and almost deterministic respectively.} of $u$ at $v \in D^*$ 
if for every $v' \in D^*$
\begin{align}
    u(v') &\ge u(v) + (v'-v) \cdot x~\qquad~\forall~v' \in D^* \label{eq:sub}
\end{align}
Let $\partial u(v)$ denote the set of all subgradients of $u$ at $v$. By IC, $x=q(v)$ satisfies (\ref{eq:sub}). Hence, $\partial u(v)$ is non-empty for any $v \in D^*$.

Define a strict linear order $\succ$ on the set of all feasible allocations in the identical-objects model, i.e., on the set $X:=\{x \in [0,1]^n: x_1 \ge \ldots \ge x_n\}$.
An example of such an ordering is the following {\bf lexicographic order}: for any $x,y \in X$, $x \succ y$ if either $x_1 > y_1$ or 
$x_1=y_1, x_2 > y_2$ or $x_i=y_i$ for $i \in \{1,2\}$, $x_3 > y_3$, or etc. If $x$ and $y$ are 
deterministic allocations (i.e., $0-1$ vectors) and $\succ$ is the lexicographic order, then $x \succ y$ 
if and only if $\sum_i x_i > \sum_i y_i$, i.e., $x$ allocates more units than $y$.

For every $v \in D^*$, let 
$$X(v):=\{x \in \partial u(v): v \cdot x \ge v \cdot y~\forall~y \in \partial u(v)\}$$
\cite{HR15} call a mechanism $(q,t)$ seller favorable if $q(v) \in X(v)$ 
for every $v$ in the domain. If $D^*$ is convex, IC implies that $u$ is convex. For convex $u$ defined 
on convex $D^*$, Rockafellar (Theorem 23.2) characterizes $v \cdot x$ for each $x \in X(v)$ (note that $v \cdot x$ 
is the same for all $x \in X(v)$) as the {\bf directional derivative} of $u$ in the direction $v$. Let $x(v)$ denote the maximal vector in $X(v)$ with respect to $\succ$.\footnote{For the proof of (iii), $x(v)$ will choose a deterministic vector and an almost deterministic vector respectively.}

Next, we define $(q^\sharp,t^\sharp)$,\footnote{Note that $(q^\sharp,t^\sharp)$ is a seller-favorable mechanism as in \cite{HR15}. In general, there exist object non-bossy mechanisms that are not seller favorable as well as seller-favorable mechanisms that are bossy.} the {\bf maximal extension} (with respect to $\succ$) of $(q,t)$  
\begin{alignat*}{1}
  q^\sharp(v) &:= x(v), \quad t^\sharp(v) := v \cdot q^\sharp(v) - u(v)~\qquad~\forall~v \in D^* 
\end{alignat*}

Note that $u^\sharp(v) = v \cdot q^\sharp - t^\sharp(v) = u(v)$ for all $v \in D^*$. As $x(v) \in \partial u(v)$, (\ref{eq:sub}) holds. Thus, $(q^\sharp,t^\sharp)$ defines an IC mechanism.
Also, $t^\sharp(v) = v \cdot x(v) - u(v) \ge v \cdot q(v) - u(v) = t(v)$ for all $v \in D^*$ as $x(v)\in X(v)$ and $q(v)\in \partial u(v)$.

To complete the proof of (i), we show that $(q^\sharp,t^\sharp)$ is object non-bossy. Take $v=(v_i,v_{-i})\in D^*$ and $ (v'_i,v_{-i}) \in D^*$. Suppose that $q^\sharp_i(v)=q^\sharp_i(v'_i,v_{-i})$. IC constraints between $v$ and $(v'_i,v_{-i})$~are
\begin{alignat*}{1}
    u^\sharp(v) - u^\sharp(v'_i,v_{-i}) &\ge (v_i-v'_i) q^\sharp_i(v_i',v_{-i}) = (v_i-v'_i) q^\sharp_i(v)\\
    u^\sharp(v'_i,v_{-i}) - u^\sharp(v) &\ge (v'_i-v_i) q^\sharp_i(v_i',v_{-i})  = (v'_i-v_i) q^\sharp_i(v) 
\end{alignat*}
\begin{align}
 \Longrightarrow \qquad   u^\sharp(v) - u^\sharp(v'_i,v_{-i}) &= (v_i-v'_i)q^\sharp_i(v)  \label{eq:e1}
\end{align}

We show that $q^\sharp(v) \in \partial u^\sharp(v'_i,v_{-i})$. Pick any $\hat{v} \in D^*$.
The IC constraint from $\hat{v}$ to $v$~is
\begin{align*}
u^\sharp(\hat{v}) &\ge u^\sharp(v) + (\hat{v}-v) \cdot q^\sharp_j(v) \\
&= u^\sharp(v'_i,v_{-i}) +  (v_i-v'_i) q^\sharp_i(v) + (\hat{v}_i - v_i)q^\sharp_i(v) + \sum_{j \ne i} (\hat{v}_j - v_j)q^\sharp_j(v)~\qquad~\textrm{(by eq. \ref{eq:e1})} \\
&= u^\sharp(v'_i,v_{-i}) + (\hat{v}_i-v'_i)q^\sharp_i(v) + \sum_{j \ne i} (\hat{v}_j - v_j)q^\sharp_j(v)
\end{align*}
Thus, $q^\sharp(v) \in \partial u^\sharp(v'_i,v_{-i})$. An identical proof shows $q^\sharp(v'_i,v_{-i}) \in \partial u^\sharp(v)$. So, $q^\sharp(v),q^\sharp(v'_i,v_{-i}) \in \partial u^\sharp(v) \cap \partial u^\sharp(v'_i,v_{-i})$. Since $(q^\sharp,t^\sharp)$ is the maximal extension of $(q,t)$ with respect to the same order $\succ$, we must have $q^\sharp(v)=q^\sharp(v'_i,v_{-i})$. This implies that $q^\sharp$ is object non-bossy. 

\noindent
(ii): If $D^*$ is convex, $u$ is convex and $\partial u(v)$ is singleton almost everywhere (Theorems 25.1 and 25.3 in Rockafellar). Consequently, 
$(q(v),t(v))=(q^\sharp(v),t^\sharp(v))$ almost everywhere.
\hfill$\blacksquare$

\subsubsection{Proofs of Section~\ref{se:rmon}}

\noindent
{\sc Proof of Proposition~\ref{pr:schur}:} We provide a proof (i) for all $v, \hat{v} \in \overline{D}^{\tinyi}$ and (ii) for all $v, \hat{v} \in D^{\tinyh}$. Thus, (\ref{eq:schur}) is satisfied for all $v,\hat{v}\in\mathcal{M}^{\tinyi}$ and almost all $v,\hat{v}\in\mathcal{M}^{\tinyh}$.

Take $v, \hat{v} \in \overline{D}^{\tinyi}$. By IC,
\begin{align}
t(\hat{v}) - t(v) &\ge \sum_{j=1}^n v_j q_j(\hat{v}) - \sum_{j=1}^n v_j q_j(v) \label{eq:wm1}
\end{align}
Let $\Delta_j(v):=v_j-v_{j+1}$ for all $j \in \{1,\ldots,n\}$,
where $v_{n+1}:=0$. As $\hat{v}, v \in \overline{D}^{\tinyi}$, $\Delta_j(v) \ge 0$ for all $j$. So,  
\begin{align}
\sum_{j=1}^n v_j q_j(\hat{v}) &= \sum_{j=1}^n q_j(\hat{v}) \Big(\sum_{k=j}^n \Delta_k(v) \Big) = \sum_{k=1}^n \Delta_k(v) \Big( \sum_{j=1}^k q_j(\hat{v})\Big) \label{eq:wm2}
\end{align}
Using (\ref{eq:wm1}) and (\ref{eq:wm2}), we have
\begin{align}
t(\hat{v}) - t(v) &\ge \sum_{k=1}^n \Delta_k(v) \sum_{j=1}^k \big( q_j(\hat{v})- q_j(v) \big) \label{eq:wm3}
\end{align}
If $q(\hat{v}) \succ_w q(v)$, then the RHS of (\ref{eq:wm3}) is non-negative. As a result, $t(\hat{v}) \ge t(v)$. This completes the proof for $(q,t)$ defined on $\overline{D}^{\tinyi}$.

Next, consider $(q,t)$ defined on domain $D^{\tinyh}$. As $(q,t)$ is symmetric, IC and IR,  Theorem~\ref{theo:sym0} implies that it is rank preserving.
 Thus, $(q,t)$ on $D^{\tinyh}$ is the symmetric extension of $(q,t)$ on $D(\sigma^{\tinyi})=D^{\tinyi}$. 
 For any $\acute{v}\in D(\acute{\sigma})$, $\grave{v}\in D(\grave{\sigma})$, (\ref{eq:sigma}) implies that $\acute{v}^{\acute{\sigma}}, \grave{v}^{\grave{\sigma}} \in D(\sigma^{\tinyi})$.  By symmetry, $t(\acute{v})=t(\acute{v}^{\acute{\sigma}})$, $t(\grave{v})= t(\grave{v}^{\grave{\sigma}})$, $q^{\acute{\sigma}}(\acute{v})=q(\acute{v}^{\acute{\sigma}})$, and  $q^{\grave{\sigma}}(\grave{v})=q(\grave{v}^{\grave{\sigma}})$.
 As weak majorization is invariant to permutations of vectors, and $(q,t)$ is symmetric,
\begin{align*}
q(\grave{v}) \succ_w q(\acute{v}) \quad \Longleftrightarrow \quad q^{\grave{\sigma}}(\grave{v}) \succ_w q^{\acute{\sigma}}(\acute{v})
 \quad \Longleftrightarrow \quad q(\grave{v}^{\grave{\sigma}}) \succ_w q(\acute{v}^{\acute{\sigma}})
\end{align*} 
Thus, the fact that (\ref{eq:schur}) for holds for types in $D(\sigma^{\tinyi})$ implies that (\ref{eq:schur}) holds for types in $D^{\tinyh}$.\hfill$\blacksquare$

\medskip
\noindent
{\sc Proof of Theorem~\ref{theo:rmon}:} \\
(a) Let $(q,t)$ be an IC and IR mechanism that satisfies majorization monotonicity 
in model~$\mathcal{M}^{\tinyi}$. By Proposition \ref{pr:nb}, there exists an object 
non-bossy mechanism $(q^\sharp(v),t^\sharp(v))$ such that 
$(q(v),t(v)) = (q^\sharp(v),t^\sharp(v))$  almost everywhere. To be precise, 
there exists a set $\check{D}^{\tinyi}\subseteq D^{\tinyi}$, where 
$D^{\tinyi} \setminus \check{D}^{\tinyi}$ has zero measure and 
$(q(v),t(v)) = (q^\sharp(v),t^\sharp(v))$ for all $v\in\check{D}^{\tinyi}$. 
Thus, $\textsc{Rev}(q,t;f^{\tinyi})=\textsc{Rev}(q^\sharp,t^\sharp;f^{\tinyi})$.  
Moreover, $(q^\sharp,t^\sharp)$ satisfies majorization monotonicity on the 
set~$\check{D}^{\tinyi}$ and is rank preserving.

Let $(v_i,v_{-i}), (\hat v_i,v_{-i}) \in \check{D}^{\tinyi}$, with $\hat v_i > v_i$. By IC, $q^\sharp_i(\hat v_i,v_{-i}) \ge q^\sharp_i(v_i,v_{-i})$. If $q^\sharp_i(\hat v_i,v_{-i})=q^\sharp_i( v_i,v_{-i})$, then object non-bossiness implies $q^\sharp(v_i,v_{-i})=q^\sharp(\hat v_i,v_{-i})$, and IC implies $t^\sharp(v_i,v_{-i})=t^\sharp(\hat v_i,v_{-i})$. 
If, instead, $q^\sharp_i(\hat v_i,v_{-i})>q^\sharp_i( v_i,v_{-i})$,  then majorization monotonicity of $(q^\sharp, t^\sharp)$ on $\check{D}^{\tinyi}$ implies that $q^\sharp(\hat v_i,v_{-i}) \succ_w q^\sharp(v_i,v_{-i})$. By Proposition~\ref{pr:schur}, we have $t^\sharp(\hat v_i,v_{-i}) \ge t^\sharp(v_i,v_{-i})$. Thus, for all $(v_i,v_{-i}), (\hat v_i,v_{-i}) \in \check{D}^{\tinyi}$, we have $\hat{v}_i > v_i$ implies $t^\sharp(\hat v_i,v_{-i}) \ge t^\sharp(v_i,v_{-i})$. This in turn implies that for all $\hat{v}, v \in \check{D}^{\tinyi}$, if $\hat{v} > v$ then $t^\sharp(\hat v) \ge t^\sharp(v)$.  Since $D^{\tinyi} \setminus \check{D}^{\tinyi}$ has zero measure, for any pair of distributions $F^{\tinyi}$ with density $f^{\tinyi}$ and $\widetilde{F}^{\tinyi}$ with density $\tilde{f}^{\tinyi}$ such that $\widetilde{F}^{\tinyi}$ first-order stochastically dominates $F^{\tinyi}$, we have
\begin{align*}
\textsc{Rev}(q,t;F^{\tinyi}) = \textsc{Rev}(q^\sharp,t^{\sharp};F^{\tinyi}) = \int_{D^{\tinyi}} t^{\sharp}(v) f^{\tinyi}(v) dv 
&\le \int_{D^{\tinyi}} t^{\sharp}(v) \tilde{f}^{\tinyi}(v) dv \\
&= \textsc{Rev}(q^\sharp,t^{\sharp};\widetilde{F}^{\tinyi})
= \textsc{Rev}(q,t;\widetilde{F}^{\tinyi})
\end{align*}
This establishes revenue monotonicity of $(q,t)$ in model~$\mathcal{M}^{\tinyi}$.

Next, let $(q,t)$ be a symmetric, IC, and IR mechanism that satisfies majorization 
monotonicity in model~$\mathcal{M}^{\tinyh}$. By Theorem \ref{theo:sym0},
its restriction to $D(\sigma^{\tinyi})$
defines an IC, IR, and rank-preserving mechanism; clearly, it satisfies majorization monotonicity.
Hence, this is a mechanism for model $\mathcal{M}^{\tinyi}$ on type space
$D^{\tinyi}$. By the argument above, there exists another IC and IR 
mechanism $(q^\sharp,t^\sharp)$ which coincides with the restriction of $(q,t)$ 
to $D(\sigma^{\tinyi})$ almost everywhere and is object non-bossy. By our earlier argument, majorization monotonicity implies that if $\hat{v} \ge v$ and $\hat{v},v \in D(\sigma^{\tinyi})$, 
then $t^\sharp(\hat{v}) \ge t^\sharp(v)$. Consequently, $t(\hat{v}) \ge t(v)$ 
for almost all $\hat{v},v \in D(\sigma^{\tinyi})$. Since $(q,t)$ is a symmetric 
mechanism, for almost all $v, \hat{v} \in \overline{D}^{\tinyh}$, with $\hat{v} 
\ge v$, we have $t(\hat{v}) \ge t(v)$.
This in turn implies revenue monotonicity of $(q,t)$ in model~$\mathcal{M}^{\tinyh}$.

(b) In model $\mathcal{M}^{\tinyi}$, let $(q,t)$ be an IC and IR mechanism which is almost deterministic. WLOG, we assume that $q$  is object non-bossy.\footnote{If $(q,t)$ is not object non-bossy then by  Proposition~\ref{pr:nb}, there exists 
another mechanism $(q^\sharp,t^\sharp)$ which is object non-bossy, almost deterministic, and rank preserving 
which agrees with $(q,t)$ except on a set of measure zero. Hence, the expected revenue from $(q^\sharp,t^\sharp)$ and $(q,t)$ is equal, and $(q,t)$ is revenue monotone if and only if $(q^\sharp,t^\sharp)$ is revenue monotone.} Let $(\hat v_i,v_{-i})$ and 
$(v_i,v_{-i})$ be two type profiles with $\hat v_i > v_i$.
By IC, $q_i(\hat v_i,v_{-i}) \ge q_i(v_i,v_{-i})$.
By object non-bossiness, if $q_i(\hat v_i,v_{-i}) = q_i(v_i,v_{-i})$,
we have $q(\hat v_i,v_{-i}) = q(v_i,v_{-i})$. 
Suppose, instead, that $q_i(\hat v_i,v_{-i}) > q_i(v_i,v_{-i})$. Since $q$ 
is almost deterministic, for all $k < i$, $q_k(\hat v_i,v_{-i}) = 1 
\ge q_k(v_i,v_{-i})$. Further, for all $k > i$, $q_k(v_i,v_{-i}) = 0
\le q_k(\hat v_i,v_{-i})$. Hence, $q(\hat{v}_i,v_{-i}) \succ_w q(v_i,v_{-i})$.
Thus, $q$ satisfies majorization monotonicity. From part (a),
$(q,t)$ is revenue monotone.

The proof for model $\mathcal{M}^{\tinyh}$ is similar to the proof in part (a).
\hfill$\blacksquare$

\medskip

\subsubsection{Proofs of Section~\ref{se:price}} 
\medskip

\noindent {\sc Proof of Theorem \ref{theo:orderedi}:}
Suppose that $(q,t)$ is an optimal deterministic (pricing) mechanism with prices $p_0=0,p_1,\ldots,p_n$. (By Proposition \ref{prop:price}, we know that a pricing mechanism representation exists for any deterministic mechanism.)

Suppose that the pricing representation of this optimal mechanism satisfies $p_1 \le \ldots \le p_n$ with at least one one strict inequality.
WLOG choose an optimal deterministic mechanism with the least number of strict inequalities in its pricing representation. The types that are allocated exactly $k$ units 
must satisfy:
\begin{align*}
\sum_{i=1}^k v_i - \sum_{i=1}^k p_i \ge \sum_{i=1}^j v_i - \sum_{i=1}^j p_i ~\qquad~\forall~j \ne k
\end{align*}
Hence, the closure of the set of types to which exactly $k$ units is sold is
\begin{align*}
R_k(p) &= \Big\{v \in D^{\tinyi}: \sum_{i=k+1}^j v_i \le \sum_{i=k+1}^j p_i~\textrm{for all~}j > k, ~ \sum_{i=j+1}^k v_i \ge \sum_{i=j+1}^k p_i~\textrm{for all}~j < k\Big\}
\end{align*}

If $v_k \ge p_k \ge p_{k-1} \ge \ldots \ge p_1$, then using $v_1 \ge v_2 \ge \ldots \ge v_k$, we have\vspace{-2mm}
\begin{align*}
\sum_{i=j+1}^k v_i \ge \sum_{i=j+1}^k p_i\qquad \forall j<k
\end{align*}
Similarly, if $v_{k+1} \le p_{k+1} \le p_{k+2} \le \ldots \le p_n$, using $v_{k+1} \ge v_{k+2} \ge \ldots \ge v_n$,
we have 
\begin{align*}
\sum_{i=k+1}^j v_i \le \sum_{i=k+1}^j p_i \qquad \forall j>k
\end{align*}
Hence,  $v \in R_k(p)$ if and only if $v_k \ge p_k$ and $v_{k+1} \le p_{k+1}$.
Thus, when $p_1 \le p_2 \le \ldots \le p_n$, we have
\begin{align*}
R_k(p) &= \big\{v \in D^{\tinyi}: v_{k+1} \le p_{k+1},\ v_k \ge p_k\big\}
\end{align*}
Consequently, the closure of the set of types where {\sl at least} $k$ units are sold is
\begin{align*}
\cup_{i=k}^n R_k(p) & = \{v \in D^{\tinyi}: v_k \ge p_k\} \\
\Longrightarrow\qquad \Pr[v\in \cup_{i=k}^n R_k(p) ] & = 1-F_k(p_k)
\end{align*}
where $F_k$ is the marginal distribution of the value of the $k^{\mbox \tiny th}$ unit.
Hence, the expected revenue from the mechanism is
\begin{align*}
\mbox{\sc Rev}(p) & = \sum_{k=1}^n p_k (1-F_k(p_k))
\end{align*}
Suppose the inequalities $p_1 \le p_2 \le \ldots \le p_n$ are strict at $j$, i.e., $p_j < p_{j+1}$. 
Consider two modifications, $\hat p$ and $\check p$, both with increasing prices, of this pricing mechanism and the expected revenue they generate: 
\begin{align*}
\hat p & = (p_0,p_1,\ldots, p_{j-1}, p_j, p_j, p_{j+2}, \ldots, p_n)\\
\mbox{\sc Rev}(\hat p) & =  \sum_{k \ne j+1} p_k (1-F_k(p_k)) + p_j (1-F_{j+1}(p_j)) \\ 
\check p & = (p_0,p_1,\ldots, p_{j-1}, p_{j+1}, p_{j+1}, p_{j+2}, \ldots, p_n)\\
\mbox{\sc Rev}(\check p) & =  \sum_{k \ne j} p_k (1-F_k(p_k)) + p_{j+1} (1-F_j(p_{j+1})) 
\end{align*}
By optimality,
\begin{align*}
\sum_{k=1}^n p_k (1-F_k(p_k)) &> \sum_{k \ne (j+1)} p_k (1-F_k(p_k)) + p_j (1-F_{j+1}(p_j)) \\
\sum_{k=1}^n p_k (1-F_k(p_k)) &> \sum_{k \ne j} p_k (1-F_k(p_k)) + p_{j+1} (1-F_j(p_{j+1}))
\end{align*}
These inequalities are strict since the new mechanisms have one less strict inequality than the original mechanism and by assumption $(q,t)$ is an optimal deterministic mechanism with the least number of strict inequalities in its pricing mechanism. The above inequalities are equivalent to
\begin{align*}
p_{j+1} (1-F_{j+1}(p_{j+1})) &> p_j (1-F_{j+1}(p_j)) \\
p_j(1-F_j(p_j)) &> p_{j+1}(1-F_j(p_{j+1})) \\[5pt]
\Longrightarrow\qquad 
\frac{1-F_{j+1}(p_{j+1})}{1-F_{j+1}(p_j)} &> \frac{p_j}{p_{j+1}} > \frac{1-F_j(p_{j+1})}{1-F_j(p_j)} \\[5pt]
\Longrightarrow\qquad \frac{1-F_j(p_j)}{1-F_{j+1}(p_j)} & > \frac{1-F_j(p_{j+1})}{1-F_{j+1}(p_{j+1})}
\end{align*}
But as $p_j < p_{j+1}$, this contradicts the assumption that $F_j$ hazard rate dominates $F_{j+1}$.
\hfill$\blacksquare$

\medskip

\noindent {\sc Proof of Theorem~\ref{theo:orderedm}:} Let $g$ be the (marginal) density of the value of each object in model $\mathcal M^{\tinyh}$. Define
	\begin{align*}
	f(v_1,v_2,\ldots, v_n) & = n!\, g(v_1)g(v_2)\ldots g(v_n)\qquad 1\ge v_1\ge v_2\ge \ldots v_n
	\end{align*}
An identical-objects model, $\mathcal{M}^{\tinyi}$, with joint density $f$ is equivalent to the  heterogeneous-objects model, $\mathcal{M}^{\tinyh}$, in the sense of Theorem~\ref{theo:sym}.\footnote{This identical-objects model is an ordered decreasing values model as defined in \citet{BM22}.} Let $F_k$ be the marginal distribution of the $k^{\mbox{\tiny th}}$ unit. By Theorem 1.B.26 in \citet{SS07},  $F_k$ hazard-rate dominates $F_{k+1}$. Hence, by Theorem~\ref{theo:orderedi}, in any optimal pricing mechanism $p^*$ if $p_1^*\le p_2^*\le \ldots\le p_n^*$ then $p_1^*=p_n^*$.
	
Let $(q^{\tinyh},t^{\tinyh})$ be an optimal deterministic symmetric mechanism for model $\mathcal{M}^{\tinyh}$ with prices $p^*_0=0, p^*_1,\ldots,p^*_n$, where the price of bundle $S$ is $\sum_{j=1}^{|S|}p^*_j$. By Theorem \ref{theo:sym}, the restriction of $(q^{\tinyh},t^{\tinyh})$ to $D(\sigma^{\tinyi})$ defines an IC and IR mechanism for model $\mathcal{M}^{\tinyi}$; denote this mechanism as $(q^{\tinyi},t^{\tinyi})$. By construction, the expected revenue of $(q^{\tinyi},t^{\tinyi})$ with density $f$ in model $\mathcal{M}^{\tinyh}$ is equal to the expected revenue of $(q^{\tinyh},t^{\tinyh})$ with i.i.d. draws from $g$ in model $\mathcal{M}^{\tinyh}$. By Theorem~\ref{theo:sym}, $(q^{\tinyi},t^{\tinyi})$ is an optimal mechanism for model $\mathcal{M}^{\tinyi}$. Also, this is a pricing mechanism with prices $p^*_0=0,p^*_1,\ldots,p^*_n$. By Theorem \ref{theo:orderedi}, there is no optimal mechanism with prices $p^*_0=0 \le p^*_1 \le \ldots \le p^*_n$ with one strict inequality. Thus, $(q^{\tinyh},t^{\tinyh})$ is not strictly supermodular.
\hfill$\blacksquare$

\subsubsection{Proofs of Section~\ref{se:dic}} 

We prove Theorem~6 for a  general domain, one that satisfies the following property.

\begin{defn}\label{de:slp} A domain of buyer types $D$ in the identical-objects model satisfies the {\bf strong-lattice property} if 
for every $v,v' \in D$, for every $k \in \{1,\ldots,n\}$ 
the types $v^\dag$ and $v^\ddag$, defined below, belong to $D$:
\begin{align}\label{eq:slp1}
v^\dag_i = 
\begin{cases}
v_i & \textrm{if}~i < k \\
\min(v_i,v'_i) & \textrm{if}~i \ge k
\end{cases}
\end{align}\vspace{-3mm}
\begin{align}\label{eq:slp2}
v^\ddag_i = 
\begin{cases}
\max(v_i,v'_i) & \textrm{if}~i \le k \\
v_i & \textrm{if}~i > k
\end{cases}
\end{align}
\end{defn}

\medskip

If (\ref{eq:slp1}) is applied for $k=1$ only and (\ref{eq:slp2}) for $k=n$ only, we get the usual lattice property.
Take two types $v$ and $v'$ in the identical-objects model and any $k=2,\ldots, n-1$.  Let $v^\dag_i $ and $v^\ddag_i$ be as defined in (\ref{eq:slp1}) and (\ref{eq:slp2}), respectively. Then $v^\dag_i $ and $v^\ddag_i $ have decreasing marginal values as  $v$ and $v'$ have decreasing marginal values. The domain $D$ satisfied the strong-lattice property if $v,\, v'\in D$ implies that  $v^\dag_i,\, v^\ddag_i\in D$. This is a richness requirement on the type space. An arbitrary type space $D$ need not satisfy this richness property but, as we show below, $\overline{D}^{\tinyi}$ and $D^{\tinyi}$ do.

 The strong-lattice property is equivalent to the lattice property under an assumption on the richness of domain of types defined next.\vspace{-3mm}
 
\begin{defn}\label{de:rich} A domain of buyer types $D$ is {\bf rich} if for every $v,v' \in D$ such that $v_k\ge v_{k+1}'$ for some $k$, then the type $v'':=(v_1,\ldots,v_k,v_{k+1}',\ldots, v_n') \in D$. \vspace{-2mm}
\end{defn}

\begin{lemma}\label{le:slp}
A rich domain of types satisfies the strong-lattice property if and only if it satisfies the lattice property.
\end{lemma}
\vspace{-3mm}

\medskip\noindent
{\sc Proof:} Suppose that the domain of types $D$ satisfies the strong-lattice property. Taking $k=1$ in Definition~\ref{de:slp}, we see that the lattice property w.r.t. with $\min$ and $\max$ is implied by the strong-lattice property. 

To prove in the other direction, take any $v, v'\in D$. The two types $v^{\dag} = \min(v,v')$ and $v^{\ddag}=\max(v,v')$ belong to~$D$ by the lattice property. First, we show $\tilde{v}$ as defined in (\ref{eq:slp1}) belongs to~$D$. Fix any $k$. If $k=1$, then $\tilde{v}=v^{\dag}$ and we are done. If $k>1$, then $\tilde{v} = (v_1,\ldots,v_{k-1},v^{\dag}_k,\ldots,v^{\dag}_n)$. By definition, $v_{k-1} \ge v_k \ge v^{\dag}_k$. Hence, richness implies that $\tilde{v} \in D$.

Next we show that $\check{v}$ as defined in (\ref{eq:slp2}) belongs to $D$. If $k=n$, then $\check{v}=v^{\ddag}$, and we are done. If $k<n$, then $\check{v}=(v^{\ddag}_1,\ldots,v^{\ddag}_k,v_{k+1},\ldots,v_n)$. As 
$v^{\ddag}_k \ge v_k \ge v_{k+1}$, richness implies that $\check{v} \in D$.
\hfill$\blacksquare$

\medskip
It is easily verified that $\overline{D}^{\tinyi}$ and $D^{\tinyi}$ are rich and satisfy the lattice property. Lemma~\ref{le:slp} implies that these domains satisfy the strong-lattice property. Using Lemma~\ref{le:slp}, one can also verify that a finite set of types defined on a  rectangular grid  satisfies the strong-lattice property.

Theorem~\ref{theo:uic} is proved below for any domain that has the strong-lattice property. 

\medskip

\noindent {\sc Proof of Theorem \ref{theo:uic}:} 
Let $(q,t)$ be a deterministic, object non-bossy, and UIC mechanism 
defined on type space $D$ that satisfies the strong-lattice property.
We start with some preliminary results. Let $k(v):=\sum \limits_{i=1}^n q_i(v),~k(v'):=\sum \limits_{i=1}^n q_i(v')$, etc.

\begin{claim}
\label{cl:2cycle}
Suppose $(q,t)$ is deterministic and UIC. Then the following are true:\vspace{-3mm}
\begin{enumerate}
\item[(i)] 
$\sum \limits_{i=1}^{k(v)} (v_i -v'_i) \ge \sum \limits_{i=1}^{k(v')} (v_i - v'_i)$
\item[(ii)] If $v \ge v'$, then
\begin{enumerate}
\item[(ii-a)] either $k(v) \ge k(v')$ or $v_i=v'_i$ for all $i \in \{k(v)+1,\ldots,k(v')\}$.
\item[(ii-b)] either $k(v) \ge k(v')$ or the IC constraints between $v$ and $v'$ bind.
\end{enumerate}
\end{enumerate}
\end{claim}

\noindent
{\sc Proof: } 
If $v \ge v'$ or $v' \ge v$, then UIC between $v$ and $v'$ implies:\begin{align}
\sum_{i=1}^{k(v)} v_i - \sum_{i=1}^{k(v')}v_i \ge t(v) - t(v') \ge \sum_{i=1}^{k(v)} v'_i - \sum_{i=1}^{k(v')}v'_i \label{eq:bic}
\end{align}
(i) Follows directly from (\ref{eq:bic}). 

\noindent
(ii-a) If $k(v') > k(v)$, then (i) implies $\sum_{i=k(v)+1}^{k(v')} (v_i-v'_i) \le 0$.
Since $v \ge v'$, this is possible only if $v_i=v'_i$ for all $i \in \{k(v)+1,\ldots,k(v')\}$.

\noindent
(ii-b) If $k(v') > k(v)$, the IC constraints in (\ref{eq:bic}) are equivalent to
\begin{align*}
\sum_{i=k(v)+1}^{k(v')}v'_i \ge t(v') - t(v) \ge \sum_{i=k(v)+1}^{k(v')}v_i
\end{align*}
Using (ii-a), the lower bound and upper bound of $t(v') - t(v)$ is the same. Hence,
the IC constraints between $v$ and $v'$ bind.
\hfill$\Box$

\begin{claim}
\label{cl:bossy}
Suppose that $(q,t)$ is a deterministic, UIC and object non-bossy mechanism. Then, the following hold.\vspace{-3mm}
\begin{enumerate}
\item[(i)] If $v \ge v'$ and $v_i=v'_i$ for all $i \le \max(k(v),k(v'))$, then, $q(v)=q(v')$
and $t(v)=t(v')$.
\item[(ii)] If $v \le v'$ and $v_i=v'_i$ for all $i > \min(k(v),k(v'))$, then, $q(v)=q(v')$ 
and $t(v)=t(v')$.
\end{enumerate}
\end{claim}

\vspace{-2mm}
\noindent
{\sc Proof:} Let $v = (v_j,v_{-j})$ and $v' = (v'_j,v_{-j})$.\footnote{If $v$ and $v'$ differ in more than one dimension, we can use this argument repeatedly.}
\vspace{-2mm}

\noindent  (i): Suppose that $v'_j < v_j$. 
Since $v_i=v'_i$ for all $i \le \max(k(v'),k(v))$, we have $j > \max(k(v'),k(v))$. Hence, $q_j(v')=q_j(v)=0$. By object non-bossiness, $q(v')=q(v)$. By UIC, we have $t(v)=t(v')$.

\vspace{-2mm}
\noindent (ii): Suppose that $v'_j > v_j$. By assumption $j \le \min(k(v),k(v'))$. Hence, $q_j(v)=q_j(v')=1$.  By object non-bossiness, $q(v)=q(v')$. By UIC, we have $t(v)=t(v')$.
\hfill$\Box$

\begin{claim}
\label{cl:dm2}
Suppose that $(q,t)$ is a deterministic, UIC and object non-bossy mechanism. Then, $q$ is increasing, i.e.,  for all $v \ge v'$, $\kappa(v) \ge \kappa(v')$.
\end{claim}

\vspace{-2mm}\noindent
{\sc Proof:} Let $v \ge v'$. Without loss of generality assume that $v' \equiv (v'_j,v_{-j})$ 
for some $j$ with $v'_j < v_j$. If $k(v') > k(v)$, then $j \notin \{k(v)+1,\ldots,k(v')\}$ 
since $v_i=v'_i$ for all $i \in \{k(v)+1,\ldots,k(v')\}$ by Claim~\ref{cl:2cycle}(ii-a). If $j \le k(v)$,
then $q_j(v)=q_j(v')=1$, and object non-bossiness implies $q(v)=q(v')$. If $j > k(v')$, then 
$q_j(v)=q_j(v')=0$, and object non-bossiness implies $q(v)=q(v')$. This gives us $k(v)=k(v')$, 
a contradiction to our assumption that $k(v') > k(v)$.
\hfill$\Box$

The next two claims prove Theorem \ref{theo:uic}.\vspace{-2mm}

\begin{claim}
\label{cl:altpr}
Suppose that $(q,t)$ is a deterministic, UIC and object non-bossy mechanism. 
Let $v$ and $v'$ be such that $v \ngeq v'$, $v' \ngeq v$ and $k(v) \ge k(v')$
Then, the incentive constraint $v'\to v$ is satisfied.
\end{claim}

\vspace{-2mm}\noindent
{\sc Proof:} We consider two cases.

\noindent {\sc Case 1:} $k(v)=n$. By Claim~\ref{cl:dm2}, the type 
$v^{\dag}=\max(v,v')$ (by the strong-lattice property, $v^{\dag} \in D$)
satisfies $k(v^{\dag}) \ge k(v)=n$. Hence, $k(v^{\dag})=n$. As $n$ units are allocated at $v$ 
and at $v^{\dag}$, and $v^{\dag}\ge v$, we have $t(v)=t(v^{\dag})$ by UIC. As UIC 
between $v'$ and $v^{\dag}$ holds, we have
\begin{align*}
\sum_{i=1}^{k(v')} v_i'-t(v') & \ge \sum_{i=1}^n v_i'-t(v^{\dag}) \\
		& =  \sum_{i=1}^{k(v)} v_i'-t(v) 
\end{align*}
Hence,  $v'\to v$ is satisfied. 

\medskip
\noindent {\sc Case 2:} $k(v) < n$. 
Define $\tilde{v}$ and $\tilde{v}'$ as follows.
\begin{align*}
\tilde{v}_i = 
\begin{cases}
v_i & \textrm{if}~i \le k(v) \\
\min(v_i,v'_i) & \textrm{if}~i > k(v)
\end{cases}
\end{align*}\vspace{-3mm}
\begin{align*}
\tilde{v}'_i = 
\begin{cases}
v'_i & \textrm{if}~i \le k(v) \\
\min(v_i,v'_i) & \textrm{if}~i > k(v)
\end{cases}
\end{align*}
By the strong-lattice property $\tilde{v},\tilde{v}' \in D$. Since $\tilde{v} \le v$, by Claim \ref{cl:dm2} we have $k(\tilde{v}) \le k(v)$. 
As $\tilde v_i=v_i$ for all $i\le k(v)=\max (k(\tilde{v}), k(v))$, by Claim 
\ref{cl:bossy}($i$), $q(\tilde{v})=q(v)$, i.e., $k(\tilde{v})=k(v)$, and
$t(\tilde{v})=t(v)$. 

Similarly, as $\tilde{v}' \le v'$, by Claim \ref{cl:dm2} we have 
$k(\tilde{v}') \le k(v') \le k(v)$. Since $\tilde{v}'_i=v'_i$ for all $i \le k(v)$ 
and $k(v') =\max(k(\tilde{v}'),k(v')) \le k(v)$, by 
Claim \ref{cl:bossy}($i$), $q(\tilde{v}')=q(v')$, i.e., $k(\tilde{v}')=k(v')$, and $t(\tilde{v}')=t(v')$.

Define a new type $\hat{v}$ as follows:\vspace{-2mm}
\begin{align*}
\hat{v}_i = \max(\tilde{v}_i,\tilde{v}'_i)~\qquad~\forall~i \in \{1,\ldots,n\}
\end{align*}
By the strong-lattice property, $\hat{v} \in D$.
Note that for all $i > k(v)$, $\hat{v}_i = \tilde{v}_i=\tilde{v}'_i = \min (v_i,v'_i)$.
Since $\hat{v} \ge \tilde{v}$, Claim \ref{cl:dm2} implies $k(\hat{v}) \ge k(\tilde v)= k(v)$.

Considering UIC from $\tilde{v}'$ to $\hat{v}$ and $\hat{v}$ to $\tilde{v}$ 
(and using $v'_i = \tilde{v}'_i$ for all $i \le k(v)$, $k(v) \ge k(v')$, and 
$t(v')=t(\tilde{v}')$, $t(\tilde{v})=t(v)$), we get
\begin{align*}
\sum_{i=1}^{k(v')}v'_i - t(v') &= \sum_{i=1}^{k(v')}\tilde{v}'_i - t(\tilde{v}') 
\ge \sum_{i=1}^{k(\hat{v})}\tilde{v}'_i - t(\hat{v}) \\
\sum_{i=1}^{k(\hat{v})}\hat{v}_i - t(\hat{v}) &\ge \sum_{i=1}^{k(v)}\hat{v}_i - t(\tilde{v}) 
= \sum_{i=1}^{k(v)}\hat{v}_i - t(v)
\end{align*}
Adding these constraints
\begin{align}
\sum_{i=1}^{k(v')}v'_i - t(v') &\ge \sum_{i=1}^{k(v)}\hat{v}_i - \sum_{i=1}^{k(\hat{v})}\hat{v}_i 
+ \sum_{i=1}^{k(\hat{v})}\tilde{v}'_i  - t(v) \label{eq:el1}
\end{align}

If $k(\hat{v})=k(v)$, then (\ref{eq:el1}) reduces to
\begin{align*}
\sum_{i=1}^{k(v')}v'_i - t(v') &\ge \sum_{i=1}^{k(\hat{v})}\tilde{v}'_i  - t(v) 
= \sum_{i=1}^{k(v)}v'_i  - t(v)
\end{align*}
where we used the fact that $v'_i=\tilde{v}'_i$ for all $i \le k(v)$. Hence, $v'\to v$ holds.

If $k(\hat{v}) > k(v)$, then (\ref{eq:el1}) reduces to
\begin{align*}
\sum_{i=1}^{k(v')}v'_i - t(v') &\ge \sum_{i=1}^{k(\hat{v})}\tilde{v}'_i - 
\sum_{i=k(v)+1}^{k(\hat{v})}\hat{v}_i  - t(v) \\
&= \sum_{i=1}^{k(\hat{v})}\tilde{v}'_i - \sum_{i=k(v)+1}^{k(\hat{v})} \tilde{v}'_i - t(v) 
\qquad~\textrm{(using $\hat{v}_i=\tilde{v}_i=\tilde{v}'_i$ for all $i > k(v)$)} \\
&= \sum_{i=1}^{k(v)}\tilde{v}'_i - t(v)  \\
&= \sum_{i=1}^{k(v)}v'_i - t(v) \qquad~\textrm{(using $v'_i=\tilde{v}_i$ for all $i \le k(v)$)}
\end{align*}
Hence, $v'\to v$ is satisfied.
\hfill$\Box$

\begin{claim}
\label{cl:altpr2}
Suppose $(q,t)$ is a deterministic, UIC and object non-bossy mechanism. Let $v$ and $v'$ be such that $v \ngeq v'$, $v' \ngeq v$ and $k(v) > k(v')$. Then, the incentive constraint $v\to v'$ is satisfied.
\end{claim}
{\sc Proof:} We consider two cases. 

\noindent {\sc Case 1:} $k(v')=0$. As $0$ units are allocated at $v'$, Claim~\ref{cl:dm2} implies that 0 units are allocated at 
$v^{\dag} = \min(v,v')$ (by the strong-lattice property $v^{\dag} \in D$). Thus, UIC implies that $t(v')=t(v^{\dag})$. 
As UIC between $v$ and $v^{\dag}$ holds, we have
\begin{align*}
\sum_{i=1}^{k(v)} v_i-t(v) & \ge -t(v^{\dag})  =  -t(v') 
\end{align*}
We conclude that $v\to v'$. 

\medskip
\noindent {\sc Case 2:} $k(v') > 0$.
Define $\tilde{v}$ and $\tilde{v}'$ as follows.
\begin{align*}
\tilde{v}_i = 
\begin{cases}
\max(v_i,v'_i)~&~\textrm{if}~i \le k(v') \\
v_i~&~\textrm{if}~i > k(v')
\end{cases}
\end{align*}
\begin{align*}
\tilde{v}'_i = 
\begin{cases}
\max(v_i,v'_i) &~\textrm{if}~i \le k(v') \\
v'_i &~\textrm{if}~i > k(v')
\end{cases}
\end{align*}
By the strong-lattice property, $\tilde{v}, \tilde{v}' \in D$. 
Since $\tilde{v} \ge v$, by Claim \ref{cl:dm2}, $k(\tilde{v}) \ge k(v) [>k(v')]$. 
Hence, $\tilde{v}_i = v_i$ for all $i > k(v')$
implies $\tilde{v}_i = v_i$ for all $i > \min(k(\tilde{v}),k(v))$. 
By Claim~\ref{cl:bossy}(ii), we have $q(v)=q(\tilde{v})$, i.e., $k(v)=k(\tilde v)$, and
$t(v)=t(\tilde{v})$. 

Similarly, $\tilde{v}' \ge v'$, and Claim \ref{cl:dm2} imply $k(\tilde{v}')\ge k(v')$. 
Hence, $\tilde{v}'_i = v'_i$ for all $i > k(v')$ implies $\tilde{v}'_i=v'_i$ for 
all $i > \min(k(\tilde{v}'),k(v'))$. By Claim \ref{cl:bossy}(ii),
we have $q(v')=q(\tilde{v}')$, i.e., $k(v')=k(\tilde v')$, and $t(v')=t(\tilde{v}')$.

Now, define $\hat{v}$ as follows:
\begin{align*}
\hat{v}_i = \min(\tilde{v}_i,\tilde{v}'_i)~\qquad~\forall~i \in \{1,\ldots,n\}
\end{align*}
By the strong-lattice property, $\hat{v} \in D$.
By Claim \ref{cl:dm2}, $q(\hat{v}) \le q(v)$ and $q(\hat{v}) \le q(v')$. 
Hence, $k(\hat{v}) \le k(v') [<k(v)]$. Applying UIC constraints
 $\tilde{v}\to\hat{v}$ and
$\hat{v}\to\tilde{v}'$, and recalling that $k(v)=k(\tilde v)$, $t(v)=t(\tilde v)$, and  $k(v')=k(\tilde v')$, $t(v')=t(\tilde v')$, we have
\begin{align*}
\sum_{i=1}^{k(\tilde v)} \tilde{v}_i - t(\tilde{v}) = \sum_{i=1}^{k(v)} \tilde{v}_i - t(v) 
&\ge \sum_{i=1}^{k(\hat{v})} \tilde{v}_i - t(\hat{v}) \\
\sum_{i=1}^{k(\hat{v})} \hat{v}_i - t(\hat{v}) &\ge 
\sum_{i=1}^{k(\tilde v')} \hat{v}_i - t(\tilde{v}') = \sum_{i=1}^{k(v')} \hat{v}_i - t(v')
\end{align*}
Adding the two constraints, we get
\begin{align}
\sum_{i=1}^{k(v)} \tilde{v}_i - t(v) &\ge \sum_{i=1}^{k(\hat{v})} \tilde{v}_i + 
\sum_{i=1}^{k(v')} \hat{v}_i - \sum_{i=1}^{k(\hat{v})} \hat{v}_i - t(v') \label{eq:al1} 
\end{align}
If $k(\hat{v})=k(v')$, then (\ref{eq:al1}) reduces to
\begin{align}
\sum_{i=1}^{k(v)} \tilde{v}_i - t(v) &\ge \sum_{i=1}^{k(v')} \tilde{v}_i - t(v') \label{eq:al2}
\end{align}
If, instead, $k(\hat{v}) < k(v')$, then (\ref{eq:al1}) reduces to
\begin{align*}
\sum_{i=1}^{k(v)} \tilde{v}_i - t(v) &\ge \sum_{i=1}^{k(\hat{v})} \tilde{v}_i 
+ \sum_{i=k(\hat{v})+1}^{k(v')} \hat{v}_i - t(v') \\
&= \sum_{i=1}^{k(\hat{v})} \tilde{v}_i + \sum_{i=k(\hat{v})+1}^{k(v')} \tilde{v}_i - t(v')
~\qquad~(\textrm{since $\tilde{v}_i=\hat{v}_i$ for all $i \le k(v')$}) \\
&= \sum_{i=1}^{k(v')} \tilde{v}_i - t(v')
\end{align*}
This is the same equation as (\ref{eq:al2}).

Next, we show that (\ref{eq:al2}) implies $v\to v'$.
Since $k(v') < k(v)$, (\ref{eq:al2}) reduces to
\begin{align*}
\sum_{i=k(v')+1}^{k(v)} \tilde{v}_i - t(v) &\ge -t(v') 
\end{align*}
Using $\tilde{v}_i=v_i$ for all $i > k(v')$ and adding 
$\sum_{i=1}^{k(v')}v_i$ on both sides, we get 
\begin{align*}
\sum_{i=1}^{k(v)} v_i - t(v) &\ge \sum_{i=1}^{k(v')}v_i-t(v') 
\end{align*}
Hence, $v\to v'$.
\hfill$\Box$

Claims \ref{cl:altpr} and \ref{cl:altpr2} prove Theorem \ref{theo:uic}. \hfill$\blacksquare$

\medskip

\noindent {\sc Proof of Theorem \ref{theo:suic}:} Let $(q,t)$ be a symmetric, rank-preserving, deterministic, 
object non-bossy, and UIC  mechanism defined on $D^{\tinyh}$. Since $(q,t)$ is rank preserving, $(q,t)$ restricted to $D(\sigma^{\tinyi})$ 
defines a feasible, object non-bossy, and UIC mechanism for the identical-objects model.
By Theorem \ref{theo:uic}, such a mechanism is IC. Hence, $(q,t)$ restricted to 
$D(\sigma^{\tinyi})$ is an IC and rank-preserving mechanism. By Theorem \ref{theo:sym0},
$(q,t)$ is an IC mechanism. 
\hfill$\blacksquare$

\bigskip

\end{appendices}





\newpage
\begin{appendices}

\section{Online Appendix}
\label{sec:onap}

This appendix, which contains some proofs and an example, is not for publication.

\noindent
{\sc Proof of Proposition~\ref{prop:sym2}:}
Suppose $(q,t)$ is an asymmetric, IC, and IR mechanism in model $\mathcal{M}^{\tinyh}$. From $(q,t)$ we construct another IC and IR mechanism $(q^*,t^*)$ which is symmetric and has the same expected revenue as $(q,t)$. Consequently, there exists an optimal mechanism which is symmetric. 

For any $\sigma\in\Sigma$, let $\sigma^{\mbox{\tiny -1}}\in \Sigma$ be such that $\sigma\sigma^{\mbox{\tiny -1}}=\sigma^{\mbox{\tiny -1}}\sigma=\sigma^{\tinyi}$. For all $v \in \overline{D}^{\tinyh}$, define
\begin{align*}
\hat{q}(v; \sigma) &:= q^{\sigma^{\mbox{\tiny -1}}}(v^\sigma) =(q_{\sigma^{\mbox{\tiny -1}}(1)}(v^\sigma),\ldots,q_{\sigma^{\mbox{\tiny -1}}(n)}(v^\sigma) ) \\
\hat{t}(v; \sigma)& := t(v^\sigma)
\end{align*}
Then for any $v, \check{v}\in D^{\tinyh}$
\begin{align*}
v\cdot \hat{q}(v; \sigma) - \hat{t}(v; \sigma)& = v\cdot q^{\sigma^{\mbox{\tiny -1}}}(v^\sigma)- t(v^\sigma)\\
		& = v^\sigma \cdot q(v^\sigma)- t(v^\sigma)\\
	& \ge  v^\sigma \cdot q(\check{v}^\sigma)- t(\check{v}^\sigma)\quad\qquad~\textrm{(since $(q,t)$ is IC)}  \\
		& =  v \cdot q^{\sigma^{\mbox{\tiny -1}}}(\check{v}^\sigma)- t(\check{v}^\sigma) \\
		& = v \cdot \hat{q}(\check{v}; \sigma)- t(\check{v}; \sigma) 
\end{align*}
Hence $(\hat q(\cdot\,; \sigma), \hat t(\cdot\,; \sigma))$ is IC. That $(\hat q(\cdot\,; \sigma), \hat t(\cdot\,; \sigma))$ is IR follows from IR of $(q,t)$. 

For all $v \in D^{\tinyh}$, define,
\begin{align*}
q^*(v) &:= \frac{1}{n!}\sum_{\sigma \in \Sigma} \hat{q}(v; \sigma)  \\
t^*(v) &:= \frac{1}{n!}\sum_{\sigma \in \Sigma} \hat{t}(v; \sigma) 
\end{align*}
The mechanism $(q^*,t^*)$ is IC and IR as it is a convex combination of IC and IR mechanisms. 
Extend $(q^*,t^*)$ to $\overline{D}^{\tinyh}$ as in Lemma \ref{le:int}.
To see that $(q^*,t^*)$ is a symmetric mechanism, note that for any fixed permutation $\check\sigma$ and any $v \in D^{\tinyh}$,\footnote{Note that even if the mechanism $(q,t)$ is deterministic (and asymmetric), the mechanism $(q^*,t^*)$ may be stochastic.}
\begin{align*}
t^*(v^{\check\sigma}) &= \frac{1}{n!}\sum_{\sigma \in \Sigma} \hat{t}(v^{\check\sigma}; \sigma) 
		 = \frac{1}{n!}\sum_{\sigma \in \Sigma} t(v^{\check\sigma\sigma}) 
		 = \frac{1}{n!}\sum_{\sigma' \in \Sigma} t(v^{\sigma'}) \\
		& =  \frac{1}{n!}\sum_{\sigma' \in \Sigma} \hat{t}(v; \sigma') 
		 = t^*(v) \\[7pt]
q^*(v^{\check\sigma}) &= \frac{1}{n!}\sum_{\sigma \in \Sigma} \hat{q}(v^{\check\sigma}; \sigma) 
		 = \frac{1}{n!}\sum_{\sigma \in \Sigma} q^{\sigma^{\mbox{\tiny -1}}}(v^{\check\sigma\sigma}) 
		= \frac{1}{n!}\sum_{\sigma \in \Sigma}q^{\check\sigma\check\sigma^{\mbox{\tiny -1}}\sigma^{\mbox{\tiny -1}}}(v^{\check\sigma\sigma})  \\[4pt]
		& = \frac{1}{n!}\sum_{\sigma' \in \Sigma}q^{\check\sigma(\sigma')^{\mbox{\tiny -1}}}(v^{\sigma'}) =  \frac{1}{n!}\sum_{\sigma' \in \Sigma} \hat{q}^{{\check\sigma}}(v; \sigma') 
		 = (q^*)^{{\check\sigma}}(v)
\end{align*}
where $\sigma'=\check\sigma\sigma$.

Finally, the expected revenue from $(q^*,t^*)$ is
\begin{align*}
\textsc{Rev}(q^*,t^*;f^{\tinyh}) &= \int \limits_{\overline{D}^{\tinyh}} t^*(v) f^{\tinyh}(v) dv = \int \limits_{\overline{D}^{\tinyh}} \frac{1}{n!}\Big(\sum_{\sigma}\hat{t}(v;\sigma)\Big) f^{\tinyh}(v) dv = \int \limits_{\overline{D}^{\tinyh}} \frac{1}{n!}\Big(\sum_{\sigma}t(v^{\sigma})f^{\tinyh}(v^{\sigma})\Big) dv \\
&= \frac{1}{n!}\sum_{\sigma}\int \limits_{\overline{D}^{\tinyh}} t(v^{\sigma})f^{\tinyh}(v^{\sigma}) dv = \frac{1}{n!}\sum_{\sigma}\int \limits_{\overline{D}^{\tinyh}} t(v)f^{\tinyh}(v) dv = \textsc{Rev}(q,t;f^{\tinyh}),
\end{align*}
where we used  exchangeability of $f^{\tinyh}$ in the third and fifth equalities. 
Hence, $(q^*,t^*)$ is a symmetric, IC and IR mechanism. By Theorem~\ref{theo:sym0}, 
$(q^*,t^*)$ is rank-preserving. Thus, for every IC and IR mechanism, 
there exists a symmetric and rank-preserving IC and IR 
mechanism that generates the same expected revenue. Hence, there exists a 
symmetric and rank-preserving optimal mechanism.\hfill$\blacksquare$

\medskip

\noindent {\sc Weak Majorization and Second-order Stochastic Dominance}

In Lemma~\ref{le:eqmaj} below, we show the equivalence between weak majorization and 
second-order stochastic dominance of allocation rules in model $\mathcal{M}^{\tinyi}$.

Let $x \equiv (x_0,x_1,x_2,\ldots,x_n=1)$ and $y \equiv (y_0,y_1,\ldots,y_n=1)$ be 
two cumulative distribution functions (cdfs) over $\{0,1,\ldots,n\}$.
Since $x$ and $y$ are cdfs, $1= x_n \ge x_{n-1} \ge \ldots \ge x_0\ge 0$ 
and $1= y_n \ge y_{n-1} \ge \ldots \ge y_0\ge 0$.

The cdf $x$ {\bf second-order stochastically dominates (SOSD)} cdf $y$, denoted $x \succ_{\tiny{SOSD}} y$, if
\begin{align*}
\sum_{i=0}^k x_i \le \sum_{i=0}^k y_i~\qquad~\forall~k \in \{0,1,\ldots,n\}
\end{align*}
Take two allocation probability vectors $q, q'\in \mathcal{M}^{\tinyi}$. The pdf over $\{0,1,\ldots,n\}$ induced 
by $q$ is
\begin{align*}
(1-q_1,\, q_1-q_2,\, q_2-q_3,\ldots, q_n)
\end{align*}
and the cdf over $\{0,1,\ldots,n\}$
\begin{align*}
F(q):= (1-q_1,\, 1-q_2,\, 1-q_3,\ldots, 1)
\end{align*}

\begin{lemma}
\label{le:eqmaj}
The following are equivalent for any pair of allocation probability vectors $q, q'\in \mathcal{M}^{\tinyi}$.
\begin{align*}
\Big[q \succ_w q'\Big] \Longleftrightarrow \Big[F(q) \succ_{\tiny{SOSD}} F(q')\Big]
\end{align*}
\end{lemma}

\noindent
{\sc Proof:}
\begin{align*}
q \succ_w q' \Leftrightarrow \sum_{i=1}^k q_k &\ge \sum_{i=1}^kq'_i\quad~\forall~k \in \{1,\ldots,n\} \\
\Longleftrightarrow \qquad\sum_{i=0}^k \big(1-F_i(q)\big) &\ge \sum_{i=0}^k\big(1-F_i(q')\big)\quad~\forall~k \in \{0,\ldots,n\} \\
\Longleftrightarrow \qquad\sum_{i=0}^k F_i(q) &\le \sum_{i=0}^kF_i(q')\quad~\forall~k \in \{0,\ldots,n\} \\
\Longleftrightarrow\qquad F(q) &\succ_{\tiny{SOSD}} F(q') \qquad\qquad\qquad\qquad\qquad\qquad\qquad\qquad \blacksquare
\end{align*}

\medskip

\noindent
{\sc Proof of Proposition~\ref{prop:price}:}
Let $(q,t)$ be a deterministic, IC, and IR mechanism. Let the range of $q$ be $R(q):=\{k(v): v \in D^{\tinyi}\}$.
By IC, if $k(v)=k(v')$, then $t(v)=t(v')$. For every $k \in R(q)$, define $P(k):=t(v)$ for some $v \in D^{\tinyi}$ with $k(v)=k$.

We now define $P$ for every $k \notin R(q)$. If $k=0 \notin R(q)$,
let $P(0):=0$. Consider $k \notin R(q)$ for some $k\ge 1$. For every $k' \in R(q)$, 
define\footnote{We use the convention that $v_0 \equiv 0$.}
\begin{align}\label{eq:ddef}
d(k',k) & := \sup_{v \in D: k(v)=k'} \Big[ \sum_{i=0}^k v_i - \sum_{i=0}^{k'}v_i\Big]\\
\label{eq:pdef}
P(k)& := \max_{k' \in R(q)}\Big[ P(k') + d(k',k) \Big]
\end{align}
This completes the definition of $P$. We now proceed in several steps. 

\noindent {\sc Step 1.} We show that $P(0)=0$. If $0 \notin R(q)$, we have $P(0)=0$ by definition.
If $0\in R(q)$, then IC implies that $q(\underline{v})=0$. Thus, $0=u(\underline{v}) = - P(0)$ implies that $P(0)=0$.

\medskip

\noindent {\sc Step 2.} We  show the IC constraints hold with respect to transfers $P$, i.e.,
\begin{align}\label{eq:ic2}
u(v) &\ge \sum_{i=0}^{k'}v_i - P(k')~\qquad~\forall v \in D^{\tinyi},~\forall~k' \in \{0,1,\ldots,n\}
\end{align}
If $k' = 0$, then the above inequality holds as $P(0)=0$ and $(q,t)$ is IR. Consider $k' \ge 1$.
If $k' \in R(q)$, then IC constraint  (\ref{eq:ic2})  follows from IC of $(q,t)$ and the definition of $P$.
Suppose, instead, that $k' \notin R(q)$. Take any $v\in D^{\tinyi}$ and suppose that $k(v)=k \in R(q)$. Then, by the definition of $P(k')$, we see that
\begin{align*}
P(k') \ge P(k) + d(k,k') \ge P(k) + \sum_{i=1}^{k'}v_i - \sum_{i=1}^k v_i \\
\Longrightarrow \qquad u(v) = \sum_{i=1}^k v_i - P(k) \ge \sum_{i=1}^{k'}v_i - P(k')
\end{align*}

\noindent {\sc Step 3.} Next, we show that $P$ is monotone. Pick $k' > k$. If $k \in R(q)$, by 
(\ref{eq:ic2}) 
for some $v \in D^{\tinyi}$ with $k(v)=k$, we have
\begin{align*}
P(k') - P(k) \ge \sum_{i=0}^{k'}v_i - \sum_{i=0}^k v_i \ge 0
\end{align*}

If $k \notin R(q)$, we consider two cases. If $k=0$, then by (\ref{eq:ic2}) for type
$\underline{v}$, $u(\underline{v})=0 \ge \sum_{i=0}^{k'}\underline{v}_i - P(k')$. Hence, $P(k') \ge 0=P(0)$, where $P(0)=0$ follows
from Step 1.

If $k \notin R(q)$ and $k \ge 1$, from (\ref{eq:pdef}) we know that there exists a $k'' \in R(q)$ 
such that 
\begin{align}
	P(k) &= P(k'') + d(k'',k) \label{eq:eee1}
\end{align}
Pick any type $v$ such that $k(v)=k''$. Then (\ref{eq:ic2}) implies 
\begin{alignat*}{1}
	\sum_{i=0}^{k''}v_i - P(k'') &\ge \sum_{i=0}^{k'}v_i - P(k') \\
	\Longleftrightarrow \quad P(k') - P(k'') &\ge \sum_{i=0}^{k'}v_i - \sum_{i=0}^{k''}v_i = \sum_{i=0}^{k'}v_i - \sum_{i=0}^{k}v_i + \sum_{i=0}^{k}v_i - \sum_{i=0}^{k''}v_i \ge \sum_{i=0}^{k}v_i - \sum_{i=0}^{k''}v_i
\end{alignat*}
where the last inequality is implied by $k' > k$. As this holds for all $v$ with $k(v)=k''$, 
we get 
\begin{alignat*}{1}
	P(k') - P(k'') &\ge \sup_{v: k(v)=k''} \Big[ \sum_{i=0}^{k}v_i - \sum_{i=0}^{k''}v_i \Big] = d(k'',k) = P(k) - P(k'')
\end{alignat*}
where the last equality follows from (\ref{eq:eee1}). Hence, we get $P(k') \ge P(k)$. Thus, $p_0:=0$ and
\begin{align*}
p_k:=P(k)-P(k-1)\ge 0\qquad \forall k\ge 1
\end{align*}
constitute a pricing mechanism that implements $(q,t)$, provided that $p_k\le\overline{v}$ for all $k\ge 1$. This is shown next.

\medskip
\noindent {\sc Step 4.} If $k \in R(q)$, then by (\ref{eq:ic2}), for some $v \in D$ with $k(v)=k$, we have \vspace{-3mm}
\begin{align*}
p_k = P(k) - P(k-1) \le \sum_{i=1}^k v_i - \sum_{i=1}^{k-1}v_i = v_k \le \overline{v}
\end{align*}
If $k \notin R(q)$, by the definition of $P(k)$, there is $k' \in R(q)$ such that
\begin{align}
P(k) = P(k') + d(k',k) \label{eq:eee2}
\end{align}
By (\ref{eq:ic2}), for any $v$ with $k(v)=k'$, we have 
\begin{align*}
\sum_{i=1}^{k'}v_i - P(k') &\ge \sum_{i=1}^{k-1}v_i - P(k-1) \\
\Longleftrightarrow \quad P(k') - P(k-1) &\le \sum_{i=1}^{k'}v_i - \sum_{i=1}^{k-1}v_i \\
&= \sum_{i=1}^{k'}v_i - \sum_{i=1}^k v_i + v_k \\
&\le \sum_{i=1}^{k'}v_i - \sum_{i=1}^k v_i + \overline{v}
\end{align*}\vspace{-3mm}
Thus, \vspace{-3mm}
\begin{alignat*}{1}
	\overline{v} + P(k-1) &\ge P(k') + \sup_{v: k(v)=k'} \Big[ \sum_{i=1}^k v_i - \sum_{i=1}^{k'}v_i\Big]
	= P(k') + d(k',k) = P(k)
\end{alignat*}
where the last equality is from (\ref{eq:eee2}). Hence, we get $p_k=P(k) - P(k-1) \le \overline{v}$.
\hfill$\blacksquare$

\begin{example}
\label{ex:uic}
{\em We present examples to show that none of the three sufficient conditions in Theorem~\ref{theo:uic}  can be dropped. In these examples, there are two identical objects with decreasing marginal values.

The type space is $\overline{D}^{\tinyi}$ for $n=2$. The mechanism is specified as follows, with buyer types identified by the number of units they receive:
\begin{align*}
(q(v),\,t(v)) & =
\begin{cases}
  ((0,0),\, 0), & \textrm{if}~v_1\le 0.5 \mbox{ and } v_1+ v_2<0.75, \qquad\qquad\quad\ \  \mbox{\bf Type 0} \\
  ((1,0),\, 0.5), & \textrm{if}~v_1> 0.5 \mbox{ and }  v_2<0.25, \qquad\qquad\qquad\qquad \mbox{\bf Type 1}  \\
  ((1,1),\, 0.75), & \textrm{if}~ v_1+ v_2>0.75 \mbox{ and }  v_2\ge 0.25, \qquad\qquad\quad\ \mbox{\bf Type 2}  \\
 ( (\tilde q,0),\, \tilde q v_1), &  \textrm{if}~ v_1+ v_2=0.75 \mbox{ and } v_1\le 0.5,\ v_2\ge 0.25 , \quad \mbox{\bf Type $\mathbf{\tilde q}$} 
  \end{cases}
\end{align*}
Type $\tilde q$ buyers are on the thick blue line-segment in Figure \ref{fig:buic}. Although each type $\tilde q$ buyer receives the same allocation, $\tilde{q} \in (0,1]$, the transfer $\tilde{q}v_1$ is a function of the buyer's value for the first unit. Thus, each type $\tilde q$ buyer has a payoff of zero.

\begin{figure}
\centering
\includegraphics[width=4in]{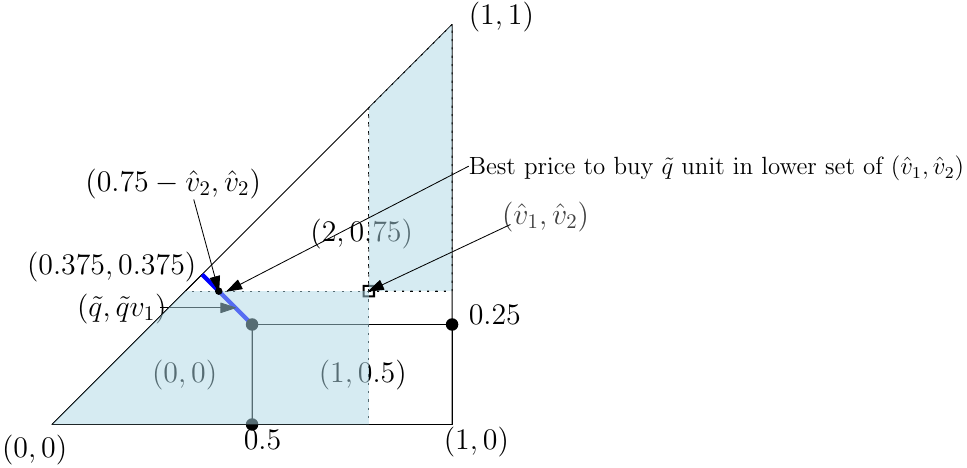}
\caption{UIC, deterministic, bossy but not IC}
\label{fig:buic}
\end{figure}

We verify that this mechanism satisfies UIC. First, note that a type 0 cannot get a positive payoff by reporting their type as $\tilde q$, 1, or 2. By UIC, type 1 buyers can misreport their type as 0 and 2 buyers only; it is easily verified that a misreport to one of these two types is not profitable. 

Next, the upper (or lower) set of any type $\tilde q$ buyer does not include any another type $\tilde q$ buyer or a type 1 buyer. Under truthful reporting, a type $\tilde q$ gets payoff zero which is the same payoff it would obtain with a misreport to a type 2 buyer, as $v_1+v_2=0.75$ for any type $\tilde q$ buyer. Its payoff remains zero by misreporting as type 0. Hence, type $\tilde q$ buyers also satisfy UIC.

Finally, consider a type 2 buyer with value, say, $(\hat{v}_1,\hat{v}_2)$. If it misreports a type $\tilde{q} $ within its UIC constraint set, it gets $\tilde{q} \in (0,1]$ units. 
The lowest possible price it pays with such a misreport is 
$\tilde{q}\max(0.375, 0.75-\hat v_2)$ (see Figure~\ref{fig:buic}). This deviation is not profitable as 
\begin{align*}
	\tilde{q}\hat{v}_1-\tilde{q}\max(0.375, 0.75-\hat{v}_2)& \le \tilde{q}[\hat{v}_1+\hat{v}_2-0.75] \\
			& \le \hat{v}_1+\hat{v}_2 - 0.75
	\end{align*}
The other UIC constraints (unprofitability of misrepresentations to a type 0 or 1) hold for a type 2 buyer.

However, any type $\tilde q$ buyer with $v_1>0.375$ profits from misreporting their type as $(0.375, 0.375)$. Thus IC, is not satisfied. In addition, IC is not satisfied for a positive measure of type 2 buyers. Take any type 2 buyer whose values satisfy $0.75<v_1+v_2\le 0.75 + \epsilon_1$, $v_1\ge 0.5-\epsilon_2$. Under truthful reporting, this buyer has a payoff less than $\epsilon_1$ in the mechanism. If this type misreports its type as $(0.375,0.375)$ its obtains a payoff of at least $\tilde q(0.125-\epsilon_2)$. This deviation is profitable when $\epsilon_1<\tilde q(0.125-\epsilon_2)$.

In this example, for any $\tilde{q} \in [0,1)$, the mechanism is UIC but not IC. If $\tilde{q} \in (0,1)$, this mechanism is object non-bossy. As the domain is $\overline{D}^{\tinyi}$, the strong-lattice property is satisfied. Hence, if $\tilde{q} \in (0,1)$, this example  shows that a stochastic, object non-bossy, and UIC mechanism in a strong lattice type space need not be IC.

If, instead, $\tilde q=1$ then the mechanism is deterministic but bossy.\footnote{While a deterministic, IC and IR mechanism is a pricing mechanism (by Proposition \ref{prop:price}), a deterministic UIC and IR mechanism need not be a pricing mechanism. In this example, all type $\tilde q$ buyers get the same allocation but their payment is type dependent; such a mechanism cannot be a pricing mechanism.}  To see this, note that the allocation to a type $\tilde q$ buyer with valuation $(v_1,v_2)$ is $q(v_1,v_2)=(1,0)$ and the allocation to a type 2 buyer with value $(v_1+\epsilon, v_2)$ is $q(v_1,v_2)=(1,1)$. Thus, in going from $(v_1,v_2)$ to $(v_1+\epsilon, v_2)$, only $v_1$ changes and only $q_2$ changes which is bossiness. 

Finally, consider another example in which the domain is the set of $\tilde{q}$ types in the previous example (the blue types in Figure \ref{fig:buic}). The domain does not satisfy the lattice property and the upper or lower set of each type is empty. Every mechanism defined on this type space is UIC. Clearly, not every mechanism is IC.

Thus, none of the assumptions of Theorem~\ref{theo:uic} can be dropped. \hfill$\blacksquare$
}
\end{example}

\end{appendices}

\end{document}